\documentclass[pre,aps,twocolumn,amsmath,amssymb,floatfix]{revtex4}
\usepackage{graphicx}
\usepackage{amsfonts}
\usepackage{epsfig}
\usepackage{dcolumn}
\usepackage{amsthm}
\usepackage{color}

\usepackage{latexsym}
\usepackage{amscd}
\usepackage{hyperref}
\usepackage[mathscr]{eucal}
 
\usepackage{bm}      
\usepackage{ifpdf}
\usepackage{bbm}
\usepackage{float}

\begin{document}

\newcommand{\newc}{\newcommand}

\newc{\beq}{\begin{equation}}
\newc{\eeq}{\begin{equation}}
\newc{\ovl}{\overline}
\newc{\bc}{\begin{center}}
\newc{\ec}{\end{center}}
\newc{\tr}{\mbox{tr}}
\newc{\pd}{\partial}
\newc{\dqv}{\delta\vec{q}}
\newc{\dpv}{\delta\vec{p}}
 \newc{\f}{\frac}
 \title{Superposition and higher-order spacing ratios in random matrix theory with application to 
complex systems}
\author{Udaysinh T. Bhosale}
\email{udaybhosale0786@gmail.com}
\affiliation{Department of Physics, National Institute of Technology, Nagpur 440010, India}
\affiliation{Physical Research Laboratory, Navrangpura, Ahmedabad 380009, India}
\date{\today}
\begin{abstract}

The statistical properties of spectra of quantum systems within the framework of random matrix theory is 
widely used in many areas of physics. These properties are affected, if two or more sets of spectra are 
superposed, resulting from the discrete symmetries present in the system. Superposition of spectra of $m$
such circular orthogonal, unitary and symplectic ensembles are studied numerically using higher-order spacing 
ratios.
%
%
For given $m$ and the Dyson index $\beta$, the modified index $\beta'$ is tabulated whose nearest neighbor 
spacing distribution is identical to that of $k$-th order spacing ratio.
%
%
For the case of $m=2$ ($m=3$) in COE (CUE) a scaling relation between $\beta'$ and $k$ is given. 
%
%
For COE, it is conjectured that for $k=m+1$ ($m\geq2$) and $k=m-3$-th ($m\geq5$) order spacing ratio distribution 
the $\beta'$ is $m+2$ and $m-4$ respectively. Whereas in the case of CSE, for 
$k=m+1$ ($m\geq2$) and $k=m-1$-th ($m\geq3$) the $\beta'$ is $2m+3$ and $2(m-2)$ respectively.
%
%
We also conjecture that for given $m$ ($k$) and $\beta$, the sequence of $\beta'$ as a function of $k$ ($m$) is 
unique. Strong numerical evidence in support of these results is 
presented. These results are tested on complex systems like the measured nuclear 
resonances, quantum chaotic kicked top and spin chains.

\end{abstract}
 
\maketitle
 
\section{Introduction}
 
Random matrix theory (RMT) has been successfully applied to study the spectral fluctuations in various complex quantum systems 
\cite{weidenmuller2009random,mitchell2010random,Haakebook,akemann2011oxford,StockmannBook,PorterBook,Forresterbook,Mehtabook,Guhr98}.
These include spin chains from condensed matter physics \cite{hutchinson2015random,wells2014quantum,rao2020higher,
rao2021critical,rao2021random}, 
nuclear physics \cite{weidenmuller2009random,mitchell2010random,gomez2011many}, chaotic billiards \cite{Bohigas84,hurt2013quantum}, 
etc. These fluctuations are used to characterize various phases of these systems, for example, thermal or localized phase of spin chains 
\cite{rao2020higher,rao2021critical,rao2021random,Oganesyan2007}, integrable to chaotic limit of the underlying classical system 
\cite{Hakke87,Reichlbook}, etc. For correct characterization of the system, its spectra needs to be desymmetrized \cite{tekur2018symmetry}. If the 
Hamiltonian $H$ for given system  an additional symmetry $S$, i.e. $[H,\hat{S}]=0$, where $\hat{S}$ is the operator corresponding to 
$S$ then the eigenvalues gets superposed. This can lead to entirely different fluctuation properties failing to characterize the 
system \cite{Dyson623,gunson1962proof,Mehtabook,anderson2010introduction,tekur2018symmetry,giraud2020probing,LeaSantos2010Localization,SantosLea12}.

Due to symmetry $S$ the $H$ becomes block diagonal in the basis formed by the eigenfunctions of $S$, i.e., 
$H=H_1 \oplus H_2 \oplus \ldots \oplus H_m$. Here, $m$ denotes the number of non-degenerate eigenvalues of $S$. 
Thus, due to the symmetry $S$, in the spectra of $H$, the eigenvalues from different blocks get superposed. 
Symmetries also have played an important role in our understanding of many areas of physics 
\cite{brading2003symmetries,ScottCorryBook,gross1996role,rosen2008symmetry}, 
mathematics \cite{tapp2011symmetry,rosen2008symmetry}, biology \cite{longo2013perspectives}, etc. 
The importance of symmetries can be understood from the works of Emmy Noether, where she has related 
continuous symmetry and conservation laws in her famous theorem \cite{sardanashvily2016noether,kosmann2011noether}.

Symmetries have also played an important role in the RMT \cite{Mehtabook,Forresterbook}. 
This goes back to Wigner who defined a class of Gaussian random matrix ensembles to understand the fluctuations 
in the nuclear spectra. The class of ensemble one uses depends on the symmetry present in the system.
In RMT, the spectral fluctuations are modeled using the most popular measure namely the nearest 
neighbour (NN) level spacings, $s_i=E_{i+1}-E_i$, where $E_i$, $i=1,2,\ldots$ are the eigenvalues of the Hamiltonian $H$.
Wigner surmised that in time-reversal invariant systems without a spin degree of freedom, these spacings are distributed as $P(s)=(\pi/2)s\exp(-\pi s^2/4)$, which indicates the level repulsion. 
For these systems, the statistical properties of the spectra are modelled correctly by the Gaussian Orthogonal 
Ensemble (GOE) having Dyson index $\beta=1$. Other ensembles that are used commonly in RMT are  
Gaussian unitary ensemble (GUE) and Gaussian symplectic ensemble (GSE) having Dyson index $\beta=2$ and $4$ 
respectively having applications to various fields \cite{akemann2011oxford,cotler2017black}.
In this work, the circular class of ensembles has been studied \cite{Forresterbook} and 
based on previous studies our results can be extended to Gaussian ensembles for large matrix dimensions under certain condition 
as explained in Sec.\ref{sec:Preliminaries}
\cite{Mehtabook,Forresterbook,tekur2018symmetry,tekurhigher2018,porter1963further,kahn1963statistical}. 
The symmetries that are used in defining respective Gaussian ensembles are the same for those of circular ensembles. 
Indices $\beta=1$, $2$ and $4$ corresponds to Dyson's threefold way and have played an important role in physics.
Matrix representation for these indices was given in the initial development of RMT.
But these ensembles are valid and exits for continuous parameter $\beta \in (0,\infty)$ and a tridiagonal 
random matrix model have been defined for them \cite{dumitriu2002matrix}. It has been used recently 
in the study of level statistics of many-body localization for $\beta \in (0,1]$ 
\cite{buijsman2018random,PiotrSierantLevelStat2019}. 
The index $\beta$ is interpreted as the inverse temperature of $T=1/\beta$ in the RMT literature.

In 1984, Bohigas, Giannoni, and Schmidt conjectured that quantum chaotic systems display NN level statistics consistent with that 
of an appropriately chosen random matrix ensemble \cite{Bohigas84,Guhr98}. 
This conjecture is supported by many theoretical studies 
\cite{Berry85,sieber2001correlations,muller2009periodic}. Due to the additional symmetry $S$, which may not be 
known apriori, the eigenvalues from different blocks get superposed. This results in level clustering of NN and 
one obtains their spacings distribution to be Poissonian 
\cite{Mehtabook,Guhr98,KarolTensor2012,KarolExtremal2013,tkocz2013note}, 
$P(s)=\exp(-s)$, which also corresponds to the NN spectral fluctuations of integrable systems known as Berry 
and Tabor's conjecture \cite{Berry77a}. This implies that to study the genuine spectral correlations, 
eigenvalues must be drawn from the same subspace.

Motivated by the works of Wigner, Dyson introduced a new class of ensembles of random matrices known as circular 
$\beta$-ensembles which are measures in the spaces of unitary matrices \cite{Dyson621}. They have played 
important roles in RMT. The Dyson index $\beta=1$, $2$ and $4$ corresponds to Circular Orthogonal Ensemble 
(COE), Circular Unitary Ensemble (CUE) and Circular Symplectic Ensemble (CSE) respectively. These ensembles have 
found applications in the scattering from a disordered cavity \cite{Forresterbook}, condensed matter and optical 
physics \cite{akemann2011oxford}. The algorithm for generating these ensembles numerically is non-trivial compared 
to that of Gaussian ensembles and is given in Ref.\cite{mezzadri2006generate}. Similar to Gaussian 
$\beta$-ensemble, the circular $\beta$-ensemble is also defined for continuous parameter $\beta\in(0,\infty)$ 
and a corresponding tridiagonal model is defined for them \cite{killip2004matrix,anderson2010introduction}.

Previous studies have shown connections within ensembles corresponding to $\beta=1$, $2$ and $4$.
A theorem which relates the properties of the CUE and COE has been conjectured in Ref.\cite{Dyson623} and 
later proved by Gunson \cite{gunson1962proof}. It states that the alternate eigenvalues obtained after 
superposition of spectra of two matrices ($m=2$ as per our notation) of the same dimension from COE belong to 
that of CUE. A similar theorem relating properties of COE and CSE were proved in Ref.\cite{MehtaDyson63}. 
It states that the alternate eigenvalues of an even-dimensional COE belong to that of CSE. Thus, these two 
theorems together state that all the statistical properties of the three ensembles are derivable from that 
of COE alone \cite{MehtaDyson63}. In fact, these two theorems hold at the level of joint probability 
distribution function (jpdf). As a corollary of these theorems, one can also say that level fluctuations of 
CUE and CSE can be obtained using COE. It is also conjectured that similar relationships hold true for 
Gaussian ensembles of infinite dimensions, in which the GOE underlies the GUE and GSE 
\cite{porter1963further,kahn1963statistical}.

 

There are recent studies where higher-order spacing ratios are studied in the superposed spectra 
\cite{tekur2018symmetry}. There it is shown that when $m$ number of COE spectra are superposed then the 
distribution of the $m-$th order spacing ratios is same as that of NN spacing ratios of the circular ensemble 
with modified Dyson index $m$.  (The definition of higher-order spacing ratios will be given in detail in 
Sec.\ref{sec:Preliminaries}). This result is then used to find symmetries in various physical systems like 
spin chains, quantum billiards and experimentally measured nuclear resonances. Similarly in 
Ref.\cite{giraud2020probing} the distribution of NN spacing as well as NN spacings ratio is studied in 
Gaussian ensembles when discrete symmetries are present with no restriction of their numbers. These results 
are then applied to quantum many-body systems, anyonic chains to periodically-driven spin systems and 
quantum clock models. It can be seen that only the special cases of spacings and 
their ratios for given $m$ are studied in Ref.\cite{tekur2018symmetry,giraud2020probing}. In this paper, 
our main aim is to study $k-$th higher-order spacing ratios for given $m$ superpositions for each of the COE, 
CUE and CSE  ensembles.
There will be no restriction on the value of $k$ as it was in Ref.\cite{tekur2018symmetry} where $k=m$. 
We will be validating our COE results by testing them on the physical model like the quantum kicked top (QKT), 
experimentally measured nuclear resonances and spin Hamiltonian.
Our results can also be used as a stringent test for studying symmetries in other systems.

The structure of the paper is as follows: In Sec.\ref{sec:Preliminaries} definition of various quantities,
namely, the NN spacing ratios, higher-order spacing ratios are given. Previous studies from random matrix theory 
and other fields using these definitions are presented. In Sec.\ref{sec:SuperpositionCOE} our results using the 
higher-order spacing ratios of superposition of COEs are presented. In Sec.\ref{sec:Testingsystems} we have 
shown the application of these results to the physical systems. In Sec.\ref{SuperpositionCUE} 
(Sec.\ref{Sec:SuperpositionCSE}) we have presented results on higher-order spacing ratios of superposition of 
CUEs (CSEs). In Sec.\ref{Sec:NumericalMethods} various numerical methods used in this paper in support of our 
results are presented. 
In Sec.\ref{Sec:Summary} summary of the results and conclusion is given.

\section{Preliminaries}
\label{sec:Preliminaries}
For the study of the spacing distribution, one needs to unfold the spectra which removes the system 
dependent spectral features, i.e., the average part of the density of states (DOS)
\cite{Mehtabook,PorterBook,Haakebook,BruusHenrik1997,Berry77a}.
This procedure is nonunique and cumbersome in many cases which can give misleading results 
\cite{GomezMisleading2002}. This difficulty can be solved by using the
NN spacing ratios \cite{VadimOganesyan2007}, i.e., $r_i=s_{i+1}/s_{i}$, $i=1$, $2$, $\ldots$,
since it is independent of the local DOS and thus does not require unfolding.
The distribution of $r_i$, $P(r)$ has been obtained for Gaussian ensembles and is given as follows 
\cite{BogomolnyDistribution2013,BogomolnyJoint2013}:
\begin{equation}
\label{Eq:PRBeta}
 P(r,\beta)=\frac{1}{Z_{\beta}} \frac{(r+r^2)^\beta}{(1+r+r^2)^{(1+3\beta/2)}},\,\,\,\,\beta=1,2,4
\end{equation}
where $Z_{\beta}$ is the normalization constant that depends on $\beta$. This quantity has found many 
applications, like numerical investigation of many-body localization 
\cite{buijsman2018random,VadimOganesyan2007,VadimOganesyan2009,ArijeetPal2010,ShankarIyer2013,cuevas2012level,biroli2012difference},
localization in constrained quantum system \cite{Anushya2018},
quantifying the distance from integrability on finite size lattices 
\cite{SantosLea10,kollath2010statistical,MarcosRigol2010,LeaSantos2010Localization,MarioCollura2012} and 
to study localization transition in L\'evy matrices \cite{TarquiniLevyMatrices}, to study symmetries
in various complex systems \cite{tekur2018symmetry,giraud2020probing}, to study degree of chaoticity in
different random matrix models \cite{AngelCorps2020}, to study quantum chaos in Sachdev-Ye-Kitaev 
models \cite{FadiSunPeriodic2020,FadiSunClassification2020,nosaka2020quantum} and 
quantum field theory \cite{srdinsek2020signatures}.
 
Variations of the spacing ratios have been studied in the recent past 
\cite{BogomolnyJoint2013,chavda2014poisson,kota2018embedded,HarshiniExact2018} including generalization to
complex eigenvalues \cite{ComplexProsen2020}.
In this work, the non-overlapping $k$-th order spacing ratio is considered, where no eigenvalue is shared 
between the spacings of numerator and denominator, defined as follows:
\begin{equation}
 r_{i}^{(k)}=\frac{s_{i+k}^{(k)}}{s_{i}^{(k)}}=\frac{E_{i+2k}-E_{i+k}}{E_{i+k}-E_{i}},\;\;\; i,k=1,2,3,\ldots
\end{equation}

This ratio has been used to study higher-order fluctuation statistics in the Gaussian \cite{Harshini2018a},
circular \cite{Harshini2018a} and 
Wishart ensembles \cite{UdaysinhBhosaleScaling2018}, and a scaling relation is given as follows:
\begin{eqnarray}
\begin{split}
P^{k}(r,\beta,m=1)&=P(r,\beta'),\,\,\,\,\beta\geq 1\\
\beta'&=\frac{k(k+1)}{2}\;\beta + (k-1),\,\,\,\,k \geq 1.
\end{split}
\label{Eq:HigherOrder}
\end{eqnarray}

\begin{table}[t!]
\begin{center}
\begin{tabular}{|c|c|c|c|c|c|c|c|c|c|}
\hline 
$k$ & $\beta=1$ & $\beta=2$ & $\beta=3$ & $\beta=4$& $\beta=5$& $\beta=6$& $\beta=7$\\ 
\hline
1&1&2&3&4&5&6&7 \\
2&4&7&10&13&16&19&22 \\
3&8&14&20&26&32&38&44 \\
4&13&23&33&43&53&63&73\\
5&19&34&49&64&79&94&109\\
6&26&47&68&89&110&131&152\\
7&34&62&90&118&146&174&202\\
8&43&79&115&151&187&223&259\\
\hline
\end{tabular}
\caption{Tabulation of higher-order indices $\beta'$ for various $k$ and $\beta$ using Eq.~(\ref{Eq:HigherOrder}).}
\label{table1}
\end{center}
\end{table}

It tells that the distribution of $k$-th order spacing ratio for a given $\beta$ ensemble is same as that 
of NN spacing ratios of $\beta'(>\beta)$ ensemble. It has been applied successfully to various physical systems 
like spin chains, chaotic billiards, Floquet systems, observed stock market, etc. 
\cite{Harshini2018a,UdaysinhBhosaleScaling2018,rao2020distribution}. It is also used recently to find the symmetries in complex 
systems \cite{tekur2018symmetry}. 

It should be noted here that the results obtained for the case $k=1$ are found to be universal since it does not depend on the 
local DOS. Whereas for $k>1$ one needs to be careful since the DOS, which changes from ensemble to ensemble, can affect the 
distribution of $r_{i}^{(k)}$. For example, in the case of circular ensembles (introduced in Sec.\ref{sec:SuperpositionCOE})
the DOS is uniform, for Gaussian ensembles it is Wigner's semicircle whereas for Wishart ensemble it is given by 
Marchenko-Pastur distribution \cite{Mehtabook}. Thus, only in the case of circular ensembles DOS will not affect the higher-order
spacing ratios. In the case of physical systems DOS can be different even if their NN fluctuation properties are explained by the same 
kind of RMT ensemble 
\cite{LeaSantos2010Localization,SantosLea12,brody81,mitchell2010random,rao2020higher,rao2021critical,rao2021random,tekurhigher2018}.
Thus, the results for $k>1$ can not be claimed to be universal that easily. But evidences from Ref.\cite{tekurhigher2018} 
(see Fig.5 therein) suggest that for given $k$ if the matrix dimension is increased large enough, which will depend on the 
RMT ensemble and the physical system under consideration, then our RMT results can be applied to them.
This means that the effects of nonuniform density can be minimised by increasing the matrix dimension
for given value of $k$. This will also be demonstrated in Sec.\ref{sec:Testingsystems} where we apply the RMT results
to the physical system of spin chain.


Recently, same relation between the higher order and the NN spacing 
distributions had been shown rigorously which is tested on random spin systems and non-trivial zeros of Riemann 
zeta function \cite{kahn1963statistical,AbulMagd1999,Katz99,Keating03,rao2020wigner,rao2020scaling}.
In Ref.\cite{tekur2018symmetry} (as explained in the Introduction)
the distribution of the $m-$th order spacing ratios after superposing the spectra of $m$ COEs is studied.
It is shown to be converging to the distribution of the NN spacing ratios $P(r,\beta')$ with $\beta'=m$ i.e.
\begin{equation}
\label{Eq:betakmresult}
P^{k}(r,1,m)=P(r,\beta'), \; \mbox{where} \; \beta'= k =  m. 
\end{equation} 

The Eq.~(\ref{Eq:HigherOrder}) is tabulated for few values of $\beta$ and $k$ in Table~\ref{table1}. It can be 
observed from the $\beta=1$ series in Table~\ref{table1} that the $\beta=4$ series appears at its even places. 
This is because of the relation between COE and CSE exists at the level of the jpdf of the eigenvalues
\cite{Dyson623,gunson1962proof} as discussed in the introduction. This observation plays an important role in 
further analysis in the subsequent part of this paper. The special case of the Eq.~(\ref{Eq:HigherOrder}) for 
$0\leq\beta\leq 1$ is given in Refs.\cite{forrester2004correlations,forrester2009random} at the level of the 
joint probability distribution of eigenvalues. There, it is shown that the jpdf of every $k$-th eigenvalue in 
certain $\beta$-ensembles with $\beta=2/k$ is equal to that of another $\beta$-ensemble with $\beta=2k$. Based 
on numerical simulations the results of our work will now be presented.

\section{Superposition and higher-order spacing ratios in COE}
\label{sec:SuperpositionCOE}

In this work, the main object of study is the circular $\beta$-ensembles. The jpdf is given as follows:
\begin{equation} 
Q_{N,\beta}[\{ \theta_i\}] = C_{N,\beta} \prod_{k>j}^{N} |\exp(i\theta_j)  -  \exp(i\theta_k) |^{\beta}
\end{equation}
where $N$ is the dimension and $C_{\beta,N}=(2\pi)^{-N}\{\Gamma(1+\beta/2)\}^{N}\{\Gamma(1+N\beta/2)\}^{-1}$ is the normalization 
constant \cite{Mehtabook,Forresterbook}. The eigenvalues $\theta_i$ are distributed uniformally on the unit circle and display level 
repulsion, characterized by $\beta\geq0$ \cite{Mehtabook}. Larger the value of $\beta$ larger is the repulsion. It can be seen that 
if $\beta=0$ is put in the jpdf all the eigenvalues are independent and thue uncorrelated. For such eigenvalues the level statistics 
follows a Poisson law.

In this section, we consider the superposition of $m\geq2$ number of COEs and study its non overlapping $k$-th order spacing 
ratio distribution $P^{k}(r,\beta,m)$. For COE, $\beta=1$ is taken and is used in modeling Hamiltonians which possess 
time-reversal symmetry and no half-integer spin \cite{Mehtabook}. This is then related with the NN spacing ratio distribution 
$P (r,\beta')$ for $\beta'>\beta$. For given $m$ and $k$ we tabulate the value of $\beta'$ for which both these distributions are 
very close to each other numerically. These values are given in Table~\ref{COEtable1} for $m=2$ to $7$ and various values of $k$.  
It can be seen that except few all values of $\beta'$ are whole number (this is the case for the superposition in CUEs and CSEs,
which will be discussed in subsequent parts of the paper). These results are plotted in Figs.~\ref{fig:COE2added}, 
\ref{fig:COE3added}, \ref{fig:COE4added}, \ref{fig:COE5added} and \ref{fig:COE6added}. The insets shows the $D(\beta')$ function, 
the minimum of which gives the best fit $P(r,\beta')$ to the numerical data $P^k(r,\beta,m)$. Its detailed definition and other 
numerical methods with which we find $\beta'$ will be discussed in Section~\ref{Sec:NumericalMethods}. We have also plotted a 
representative data for noninteger $\beta'$ in Sec.\ref{SuperpositionCUE} (see Fig.\ref{fig:CUE3added3}) where we have studied the 
superposition of CUEs.

The $m=2$ is an interesting case for which we have obtained scaling relations for even and odd values of 
$k$ and $\beta'$ as given below:
\begin{equation}
\label{Eq:COEseries1}
 \beta'=\frac{5k}{2}-3+\frac{(k-2)(k-4)}{4},\,\,\,k=2,4,6,\ldots
\end{equation}
and
\begin{equation}
\label{Eq:COEseries2}
 \beta'=3k-5+\frac{(k-3)(k-5)}{4},\,\,\,k=3,5,7,\ldots
\end{equation}
For even $k$ the scaling relation reduces to Eq.~(\ref{Eq:HigherOrder}) for $\beta=2$ by suitable change of 
variables as $l=k/2$ where $l=1,2,3,\ldots$. This is a known result on the connection between CUE and 
superposition of two COEs \cite{Dyson623,gunson1962proof}. For odd $k$ changing the variable as $q=(k-1)/2$ 
then Eq.~\ref{Eq:COEseries2} reduces to a simpler form as $\beta'=q(q+3)$ for $q=1,2,3\ldots$. It can be 
compared with Eq.~(\ref{Eq:HigherOrder}) and can be seen that it does not reduces to Eq.~(\ref{Eq:HigherOrder}) 
for any $\beta$. Thus, using the scaling relation in Eq.~(\ref{Eq:COEseries2}), no statement can be made at 
the level of jpdf of circular $\beta$-ensemble.

Based on the results in Table~\ref{COEtable1}, we have given two conjectures on the lines of 
Eq.(\ref{Eq:betakmresult}) 
(see Ref.\cite{tekur2018symmetry}) at the level of spectral fluctuations. The first conjecture is as follows:
\begin{equation}\label{Eq:Conjecture2}
P^{k}(r,1,m)=P(r,\beta'), \; \mbox{for} \; \beta'=k+1=m+2 
\end{equation}
and $m\geq2$, while the second one is as follows:
\begin{equation}\label{Eq:Conjecture3}
P^{k}(r,1,m)=P(r,\beta'), \; \mbox{for} \; \beta'=k-1=m-4 
\end{equation}
and $m\geq5$. Our conjectures hold true for asymptotic value of $N$. These conjectures have appeared in our arXiv
preprint \cite{bhosale2019superposition}.

\begin{table}[t!]
\begin{center}
\begin{tabular}{|c|c|c|c|c|c|c|}
\hline 
$k$ & $m=2$ & $m=3$ & $m=4$ & $m=5$ & $m=6$& $m=7$\\
&$\beta'$&$\beta'$&$\beta'$&$\beta'$&$\beta'$&$\beta'$\\ 
\hline
1&0&0&0&0&0&0\\
2&2&1.25&1&1&0.75&0.75\\
3&4&3   &2.5&2&2&2\\
4&7&5&4 &3.5&3&3\\
5&10&7.5&6&5&5&4\\
6&14&10 &8&7&6&6\\
7&18&13 &11&9&8&7\\
8&23&17 &13&12&10&9\\
9&28&20 &16&14&12&11\\
10&34&24&20&17&15&13\\
11&40&29&23&20&17&16\\
12&47&33&27&23&20&18\\
13&54&38&31&26&23&21\\
14&  &43&35&30&26&23\\
15&  &49&39&33&29&26\\
16&  &  &44&  &32&29\\
17&  &  & &  & &32\\
18&  &  & &  & &35\\
19&  &  & &  & &39\\
20&  &  & &  & &42\\
\hline
\end{tabular}
\caption{Tabulation of higher-order indices $\beta'$ for various $k$ and superposition of $m$ COEs each 
having dimension $N=8400$.}
\label{COEtable1}
\end{center}
\end{table}

\begin{figure}[h!]
\begin{center}
\includegraphics*[scale=0.35]{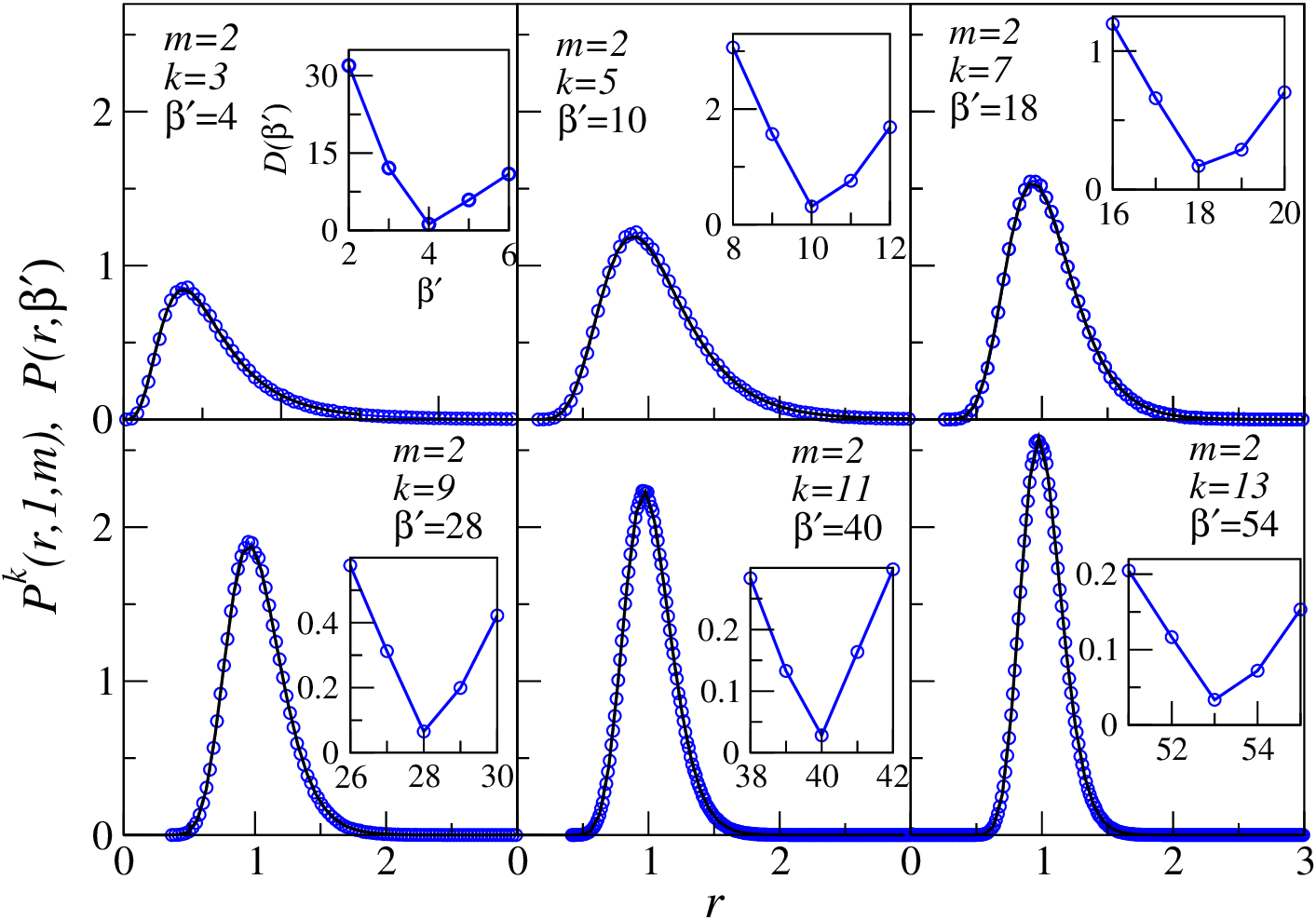}
\caption{(Color online) Distribution of the $k$-th order spacing ratios (circles) for a superposition of 
$m=2$ COE spectra. The dimension of the matrices is $N=8400$. The solid curve corresponds to 
$P(r,\beta')$ as given in Eq.(\ref{Eq:PRBeta}) with $\beta'$ given in Table~\ref{COEtable1}. 
The insets shows $D$ as a function of $\beta'$.}
\label{fig:COE2added}
\end{center}
\end{figure}
\begin{figure}[h!]
\begin{center}
\includegraphics*[scale=0.35]{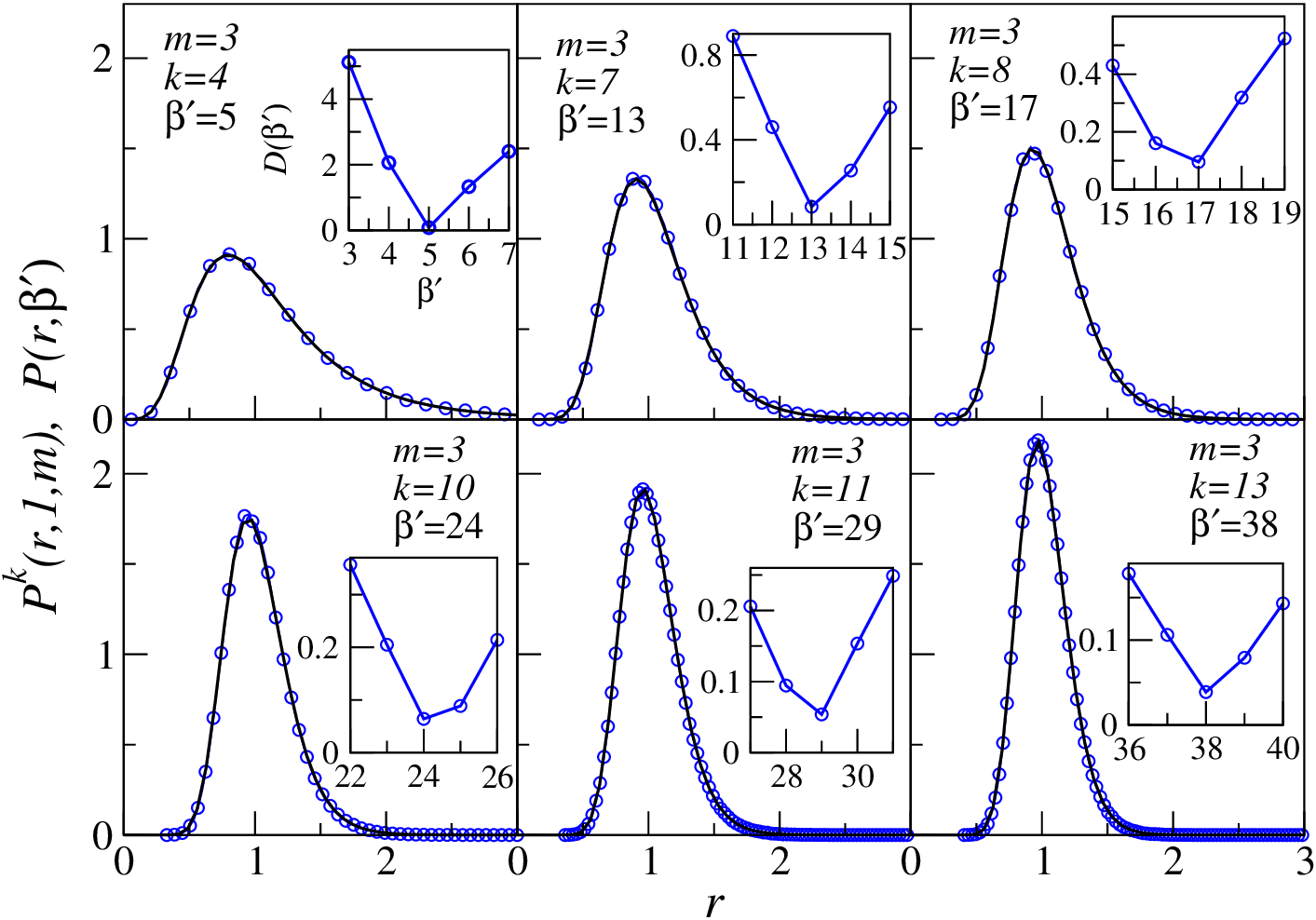}
\caption{(Color online) Same as Fig.\ref{fig:COE2added} for $m=3$.}
\label{fig:COE3added}
\end{center}
\end{figure}
\begin{figure}[h!]
\begin{center}
\includegraphics*[scale=0.35]{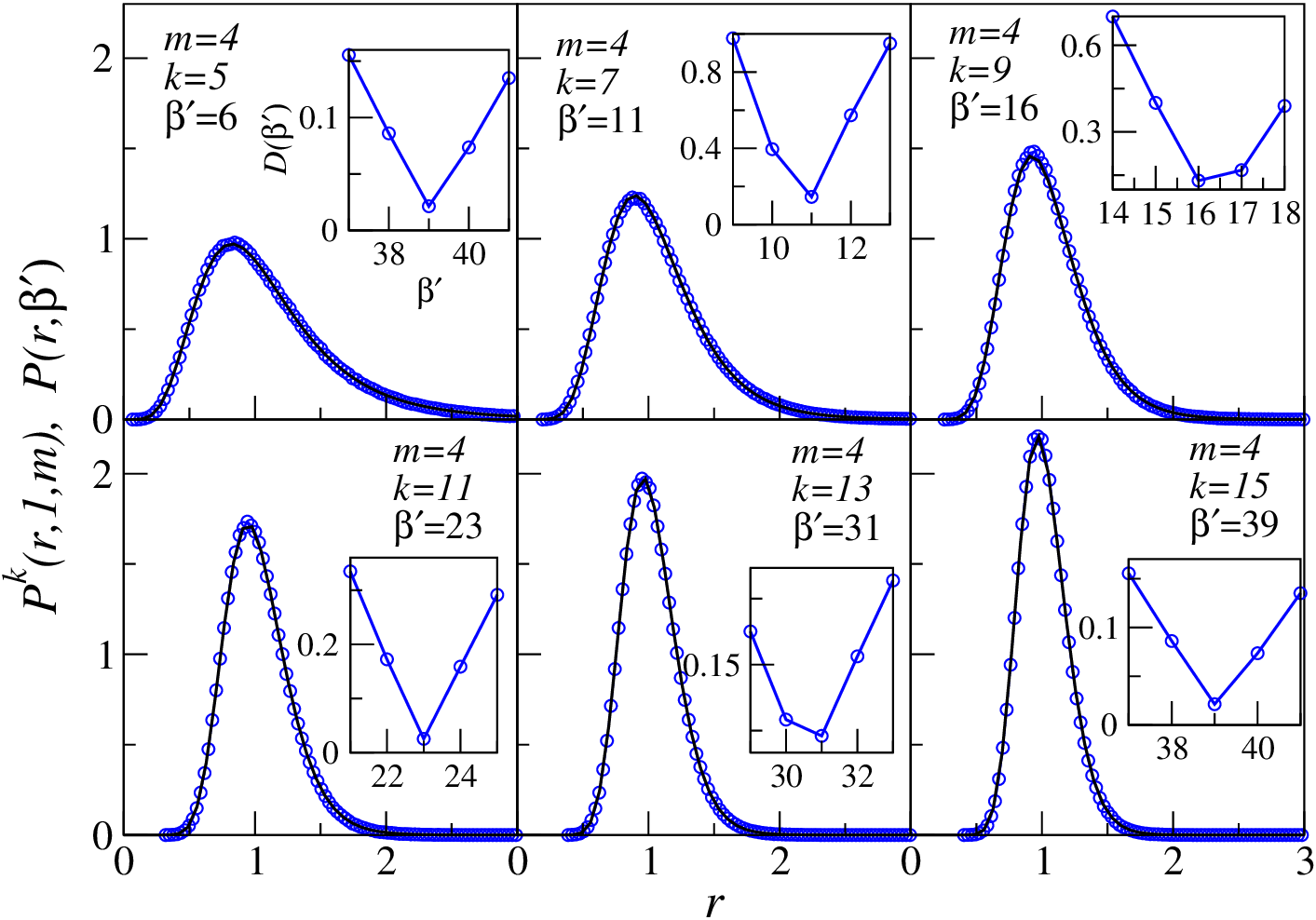}
\caption{(Color online) Same as Fig.\ref{fig:COE2added} for $m=4$.}
\label{fig:COE4added}
\end{center}
\end{figure}
\begin{figure}[h!]
\begin{center}
\includegraphics*[scale=0.35]{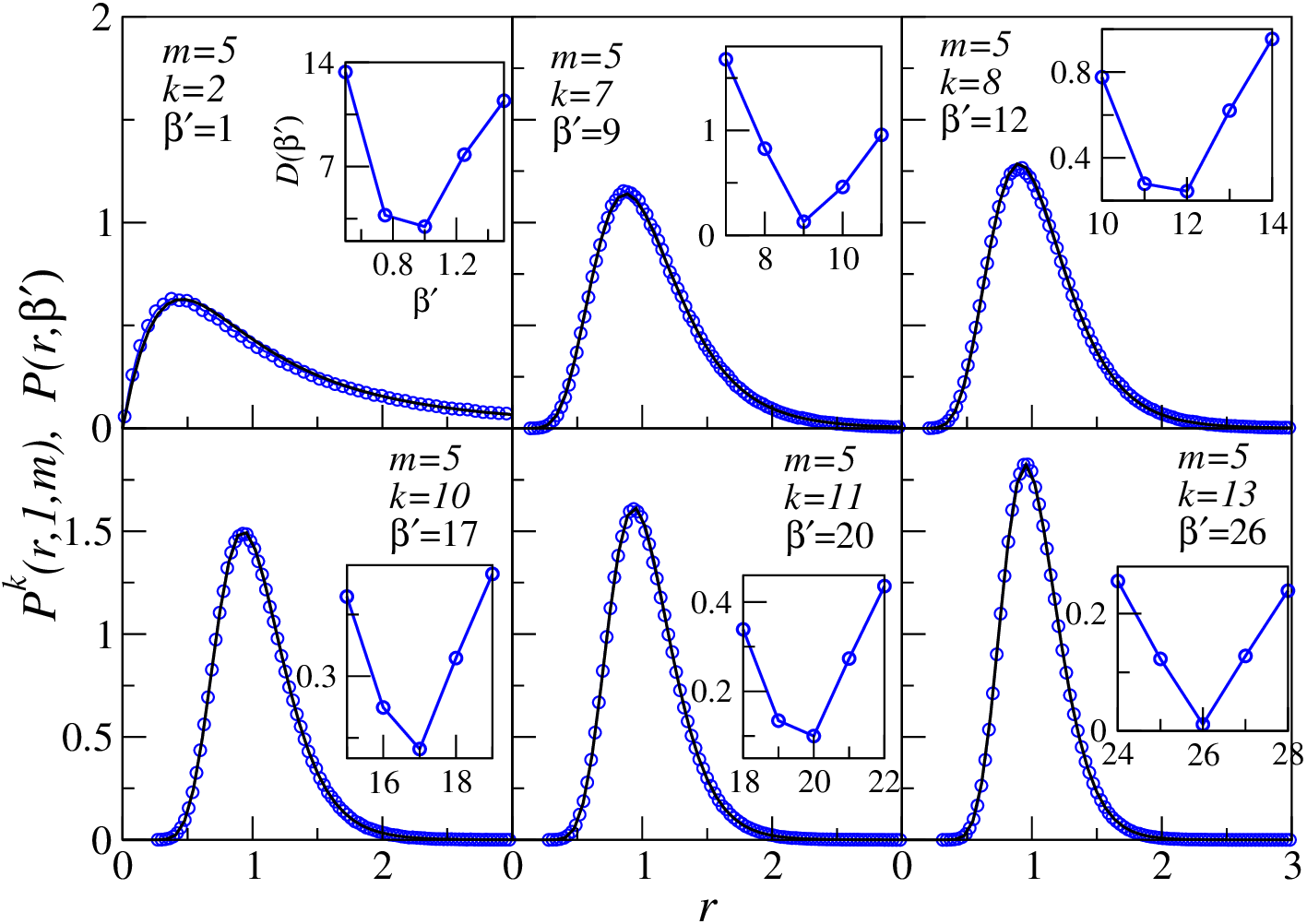}
\caption{(Color online) Same as Fig.\ref{fig:COE2added} for $m=5$.}
\label{fig:COE5added}
\end{center}
\end{figure}
\begin{figure}[h!]
\begin{center}
\includegraphics*[scale=0.35]{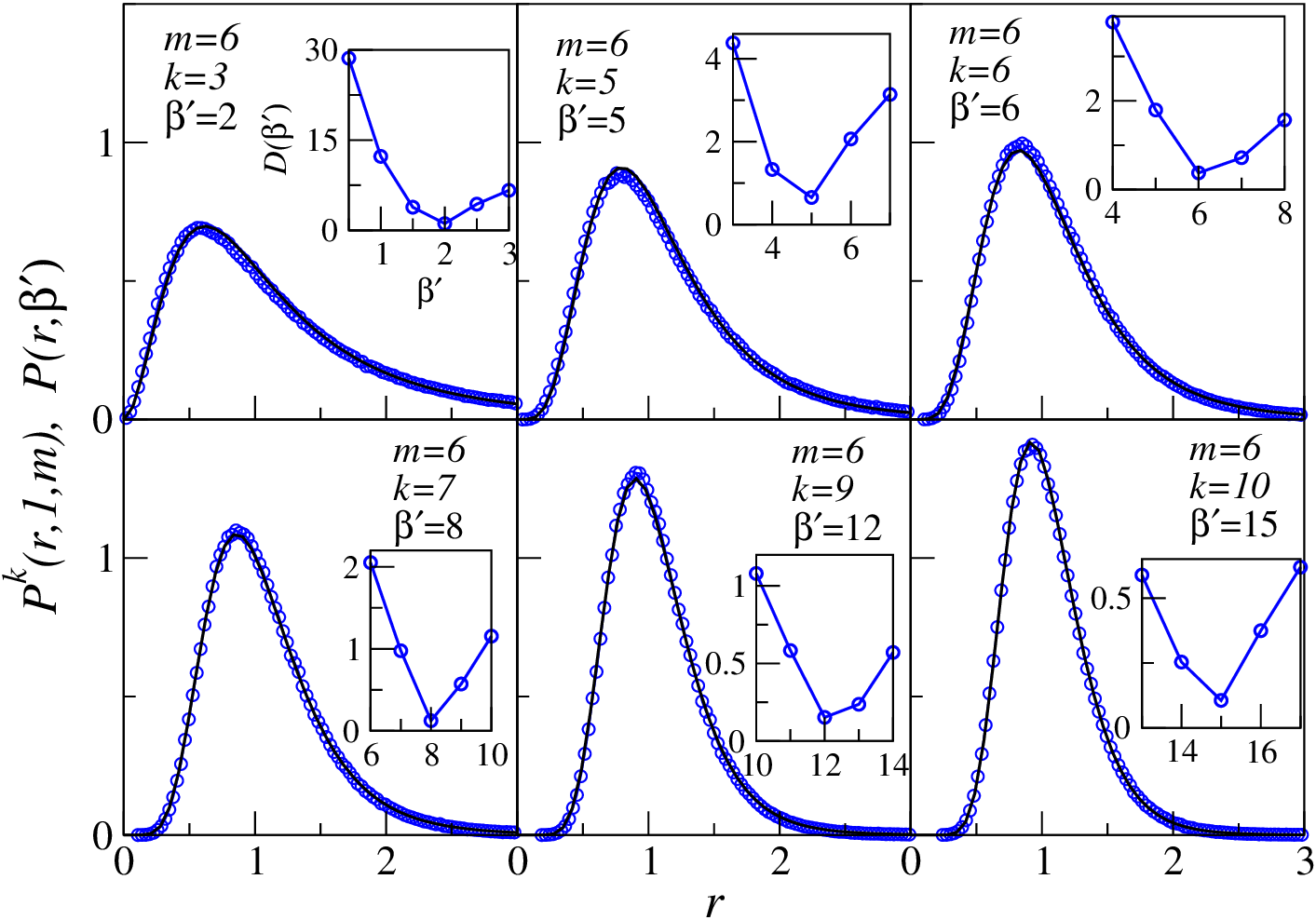}
\caption{(Color online) Same as Fig.\ref{fig:COE2added} for $m=6$.}
\label{fig:COE6added}
\end{center}
\end{figure}

\section{Testing RMT results to physical systems}
\label{sec:Testingsystems}

In this section, we test our COE results on the physical model of the quantum kicked top (QKT), experimentally measured 
data of nuclear resonances and a spin Hamiltonian. Using these systems it will be shown that our results hold true for $m=2$ case. 
First, we consider the model of QKT. This is a fundamental and important time-dependent 
model for the chaotic Hamiltonian system. This model was introduced in \cite{Hakke87} and has been the topic
of theoretical and experimental research since then 
\cite{Hakke87,JakubZakrzewski1991,RobertAlicki1996,SethLloyd2002,MarekKus04,Chaudhury09,Lombardi2011,
ZbigniewPuchala2016,Neill2016,Udaysinhbifurcation2017,Arjendu2017,UdaysinhPeriodicity2018,
KrithikaKickedTop2019,EricMeier2019,DeutschFeedback2020,TianruiXu2020,SantoshKumar2020}.
It has been realized in various experiments, namely, the hyperfine states of cold atoms \cite{Chaudhury09},
three coupled superconducting qubits \cite{Neill2016} and in a two-qubit NMR system 
\cite{KrithikaKickedTop2019}. This is also an important model from the perspective of random matrix theory 
and quantum information. This model shows regular to chaotic behavior as a function of chaoticity parameter. 
Effects of the underlying phase space on various measures quantum correlations are studied 
\cite{Miller99,Neill2016,Udaysinhbifurcation2017,Arjendu2017}.  
For classical limit being fully chaotic, the NN fluctuations of symmetry reduced spectra of the QKT 
corresponds to that of COE ensemble \cite{Hakke87,Bohigas84}.

QKT is characterized by an angular momentum vector ${\bf{J}} = (J_x, J_y, J_z)$ and its components obey
the standard algebra of angular momentum. The unitary time evolution operator for QKT is given as follows:
\begin{equation}
 \widehat{U}=\exp\left(-ipJ_y \right) \exp\left(-i\dfrac{k}{2j}J_z^2 \right) 
\end{equation}
It represents free precession of the top around $y$ axis with angular frequency $p$ while the second 
term is periodic $\delta$ kicks applied to the top. Here, $k$ is called as the kick strength or chaos 
parameter. For $k = 0$ the top is integrable and for $k > 0$ it becomes increasingly chaotic. 
For given $j$ the Hilbert space dimension is equal to $2j+1$. 

As discussed in Ref.\cite{Hakke87}, for $p\neq\pi/2$, the case relevant for us, there are two symmetries 
present in QKT, since $\widehat{U}$ commutes with the rotation operator $\hat{R}_y$ having two eigenvalues. 
Thus, the matrix representation of $\widehat{U}$ in the basis of $R_y$ is block diagonal consisting of two 
blocks and their dimensions are $j$ and $j+1$. The large $j$ case is relevant of us as these dimensions are 
very close to each other. For the fully chaotic case, the eigenvalue fluctuations of 
$\widehat{U}$ in each such block is found to follow COE statistics \cite{Hakke87}. Taking these eigenvalues 
together and studying their higher-order spacings ratio is an ideal case for our study. We can compare them 
with our COE results of $m=2$.

For our study $j=1000$ and twenty realizations for large and different values of $k$ is taken. Thus, 
the dimension of the matrix $\widehat{U}$ is $2001$. The dimension of two blocks when $\widehat{U}$ is written 
in the 
eigenbasis of $\hat{R}_y$ is $1000$ and $1001$ respectively. Thus, we can test our RMT results of $m=2$
case of COE. The results are plotted in Figs.~\ref{fig:Kickedtop1} and  \ref{fig:Kickedtop2} for $k=2$ to 
$13$. 
It can be seen here that the results agree very well with the RMT results, $m=2$ case of COE in
Table~\ref{COEtable1}, thus implying that there are two symmetries in the QKT which we were already known 
apriori \cite{Hakke87}. The RMT results hold true for such a large value of $k$ due to the uniform density in both, the COE ensemble
and the QKT (as discussed in the Sec.\ref{sec:Preliminaries}).

 

Now, we go on to test our results to experimentally measured nuclear resonances of Tantalum (Ta$^{181}$) 
\cite{HackenNeutron1978,brody81}. It is known that it belongs to the GOE and there are two symmetries present in 
it \cite{HackenNeutron1978,brody81}. We have $434$ such resonances and the results are plotted in 
Fig.(\ref{fig:Nucleardata1}). The value of $\beta'$ is chosen such that $P(r,\beta')$ is best fit to the 
data. The results are compared with $m=2$ case of COE in Table~\ref{COEtable1}.
It can be seen that the results 
hold true only for $k=2$ and $3$. From $k=4$ onwards we observe deviations from our RMT results.
This is due to the nonuniversal effects in the DOS (not shown here) and the small sample size. 
With this example, we have tested our COE result on a GOE system with 
small sample size and found the number of symmetries in it successfully.

Now, a Hamiltonian corresponding to spin-1/2 chain \cite{SantosLea12,TheodoreHsu1993} is considered as follows:
\begin{equation}
\begin{split}
H=&\sum_{i=1}^{L-1}\left[J_{xy} \left(S_i^x S_{i+1}^x + S_i^y S_{i+1}^y\right) +J_z S_i^z S_{i+1}^z\right]\\
&+\alpha \sum_{i=1}^{L-2}  \left[J_{xy}' \left(S_i^x S_{i+2}^x + S_i^y S_{i+2}^y\right) +J_z' S_i^z S_{i+2}^z\right]
\end{split}
\end{equation}
where $L$ is total number of sites, the NN coupling strengths in three directions are denoted by $J_{xy}$ and $J_{z}$ 
(couplings in $x$ and $y$ directions are same). Similarly, $J_{xy}'$ and $J_z'$ are the next-NN coupling strengths. For $\alpha=0$ 
this Hamiltonian is integrable \cite{tekur2018symmetry}. Whereas it is chaotic for $\alpha \gtrsim 0.2$ and follows
GOE statistics \cite{SantosLea12}. There are various symmetries 
in this model \cite{santos2009transport,kudo2005level}. The first one is due to conservation of total spin in the $z$ direction 
denoted as $S_z=\sum_{i=1}^{L} S_i^z$. For our work we are restricting to the case $S^z=0$ (even $L$) and  $S^z=-1/2$ (odd $L$) such 
that the block-matrix is of maximum possible dimension. The Hamiltonian commutes with the parity operator with eigenvalues $\pm 1$ 
leading to two invariant subspaces in a given $S_z$ block. Results for this case are plotted in Fig.\ref{fig:SpinChain2Sym1} for odd 
value of $L$. It can be seen that our RMT results for $m=2$ case from Table\ref{COEtable1} holds true for $k=2$, $3$ and $4$. 
For even $L$, the Hamiltonian also commutes with the operator corresponding to rotation symmetry 
with eigenvalues $\pm 1$. Thus, in this case there will be totally four invariant subspaces in a given $S_z$ block. The results for 
this case are plotted in Figs.\ref{fig:SpinChain4Sym1} and \ref{fig:SpinChain4Sym2}. In this case our results hold from $k=2$ to 
$6$ with the corresponding RMT results for $m=4$ from Table\ref{COEtable1}. Thus, our $m=2$ and $m=4$ the COE results agree with spin 
Hamiltonian having GOE statistics when there are two and four symmetries are present in it respectively. In the later case the block-matrix dimension is 
$3432$ whereas in the previous case it is $1716$. 
In this case also the deviations are due to the nonuniversal effects in DOS which is Gaussian in nature (not shown here).
But the results seems to improve due to the increased matrix dimension.




\begin{figure}[h!]
\begin{center}
\includegraphics*[scale=0.35]{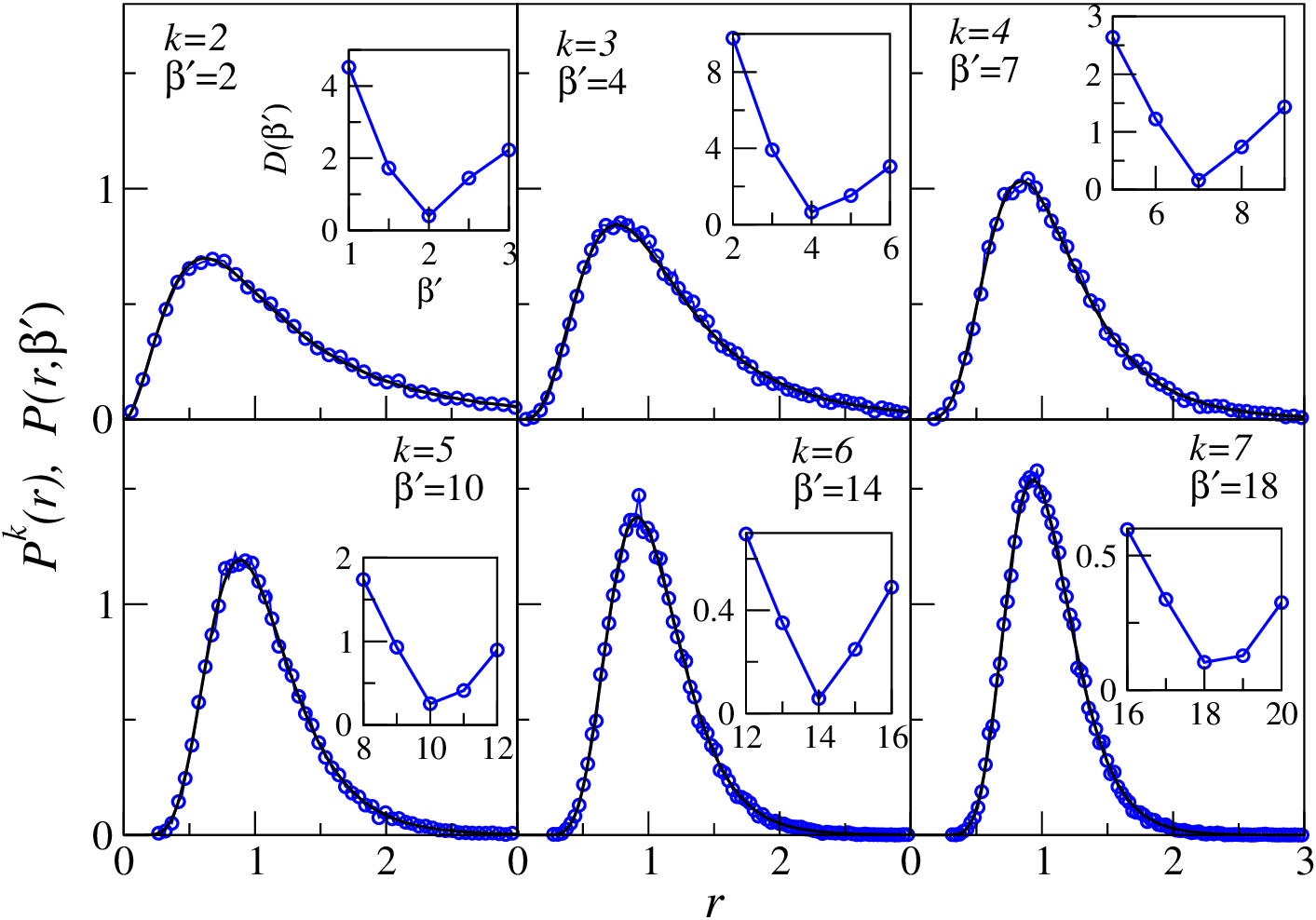}
\caption{(Color online) The distribution of the $k$-th order spacing ratios for $k=2$ to $7$ is shown for the QKT.
The numerical data $P^k(r)$ (circles) are obtained from the computed eigenvalues of QKT. The solid line represents 
$P(r,\beta'),$ with $\beta' = 2,4,7,10,14,18$.}
\label{fig:Kickedtop1}
\end{center}
\end{figure}
\begin{figure}[h!]
\begin{center}
\includegraphics*[scale=0.35]{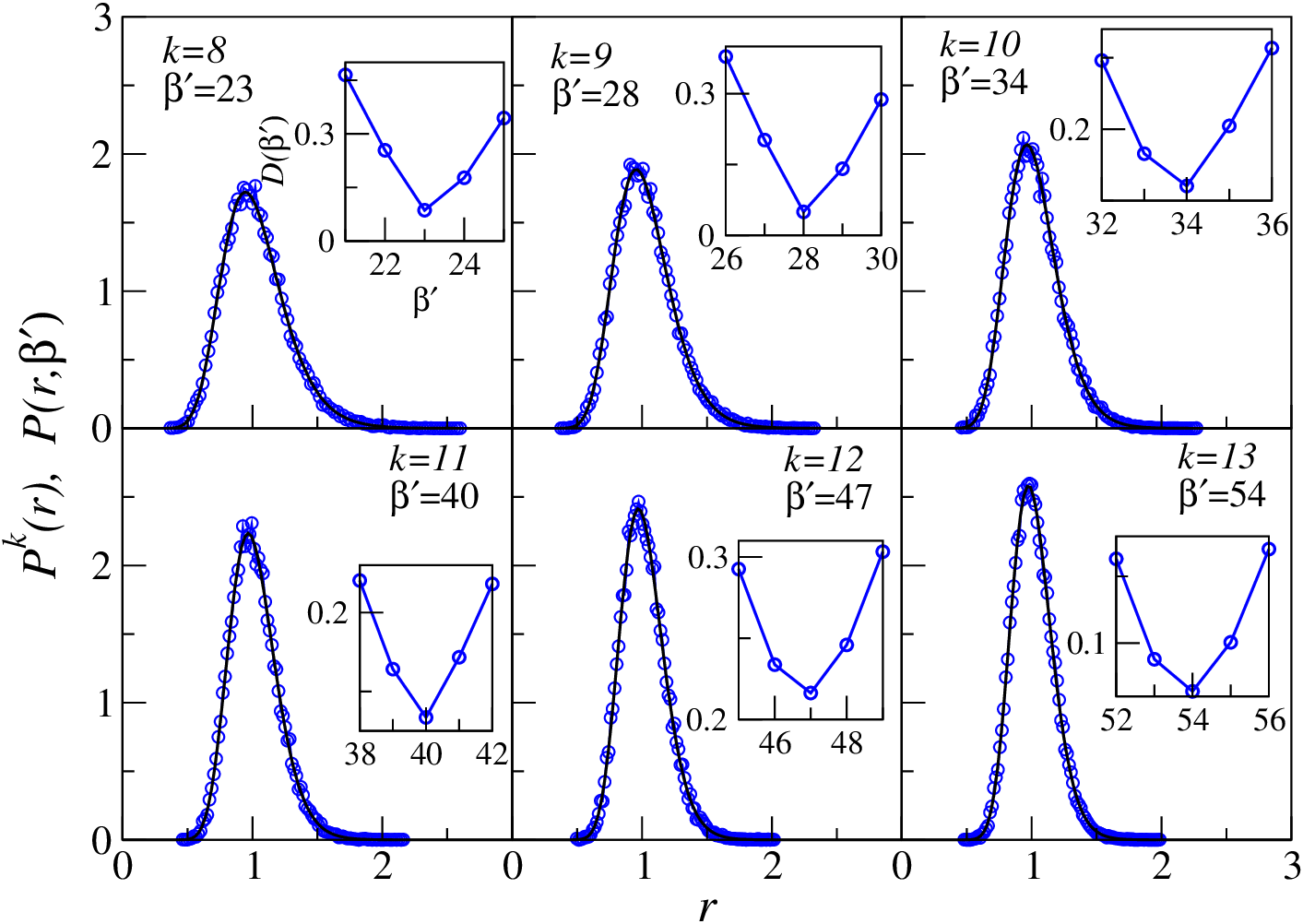}
\caption{(Color online) Same as Fig.~\ref{fig:Kickedtop1} but for $k=8$ to $13$ and $\beta'=23,28,34,40,47,54$.}
\label{fig:Kickedtop2}
\end{center}
\end{figure}
\begin{figure}[h!]
\begin{center}
\includegraphics*[scale=0.35]{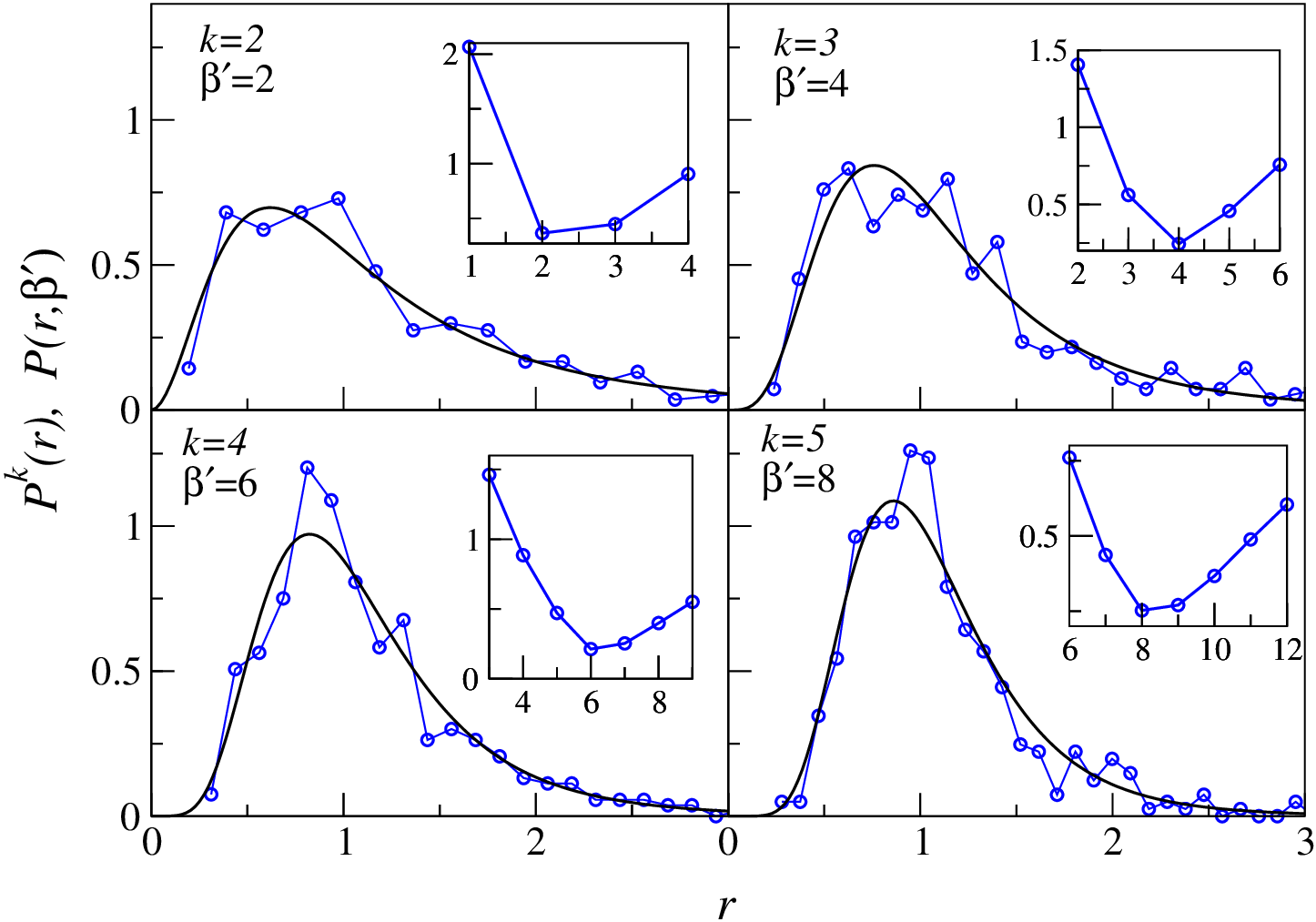}
\caption{(Color online) Same as Fig.~\ref{fig:Kickedtop1} but for experimentally measured nuclear resonances
of Ta$^{181}$. Here, $k$ varies from $2$ to $5$ while for solid curves $\beta'=2,4,6$ and $8$.}
\label{fig:Nucleardata1}
\end{center}
\end{figure}

\begin{figure}[h!]
\begin{center}
\includegraphics*[scale=0.35]{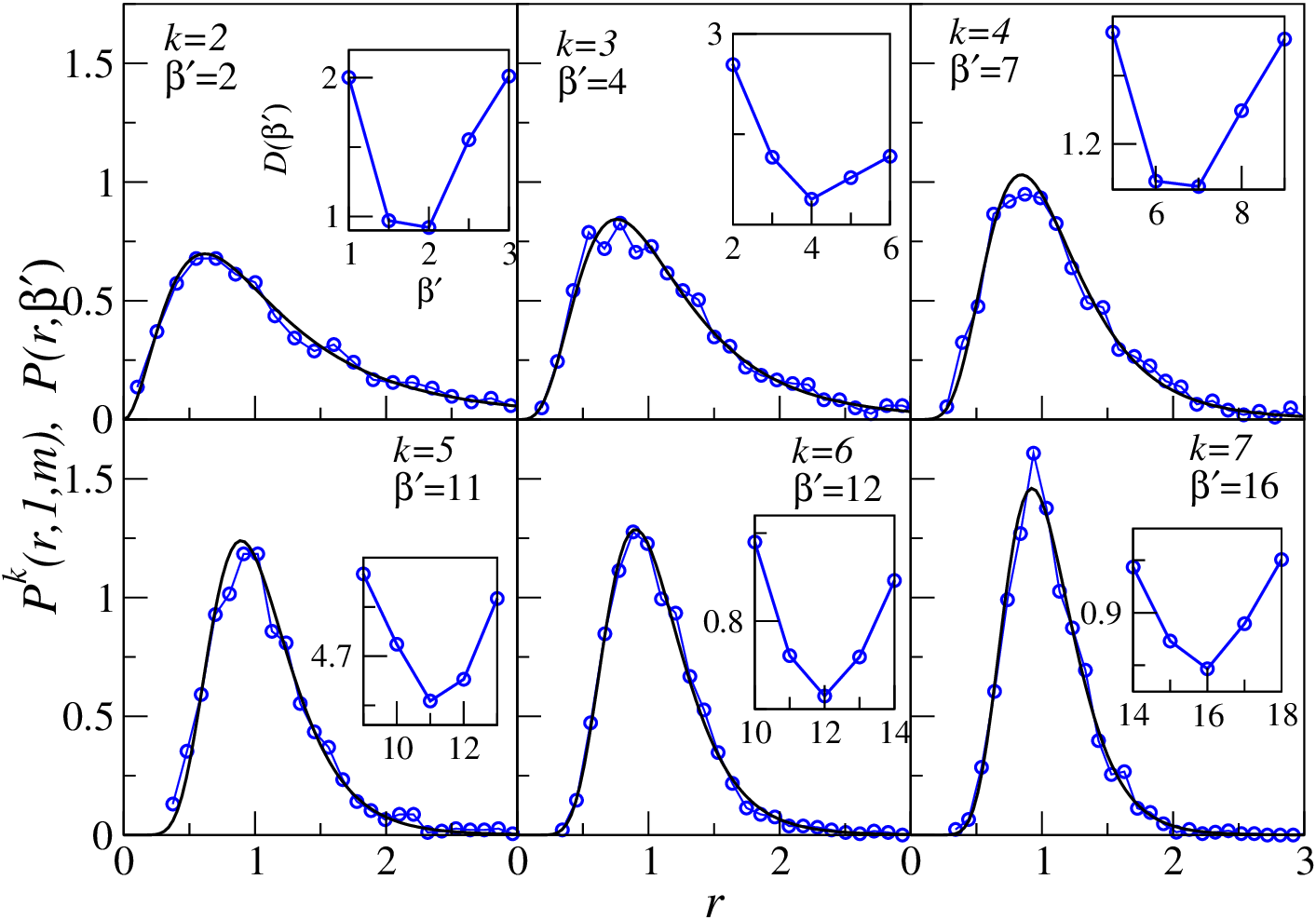}
\caption{(Color online) The distribution of the $k$-th order spacing ratios for $k=2$ to $7$ is shown for the spin-1/2 chain 
Hamiltonian with $L=13$ with $7$ up spins, $J_{xy}=J_{xy}'=1$, $J_z=J_z'=0.5$ and $\alpha=0.5$. The dimension of the block-matrix is 
$1716$. The numerical data $P^k(r)$ (circles) are obtained from the computed eigenvalues of the Hamiltonian. The solid line 
represents $P(r,\beta'),$ with $\beta' = 2,4,7,11,12,16$.}
\label{fig:SpinChain2Sym1}
\end{center}
\end{figure}
\begin{figure}[h!]
\begin{center}
\includegraphics*[scale=0.35]{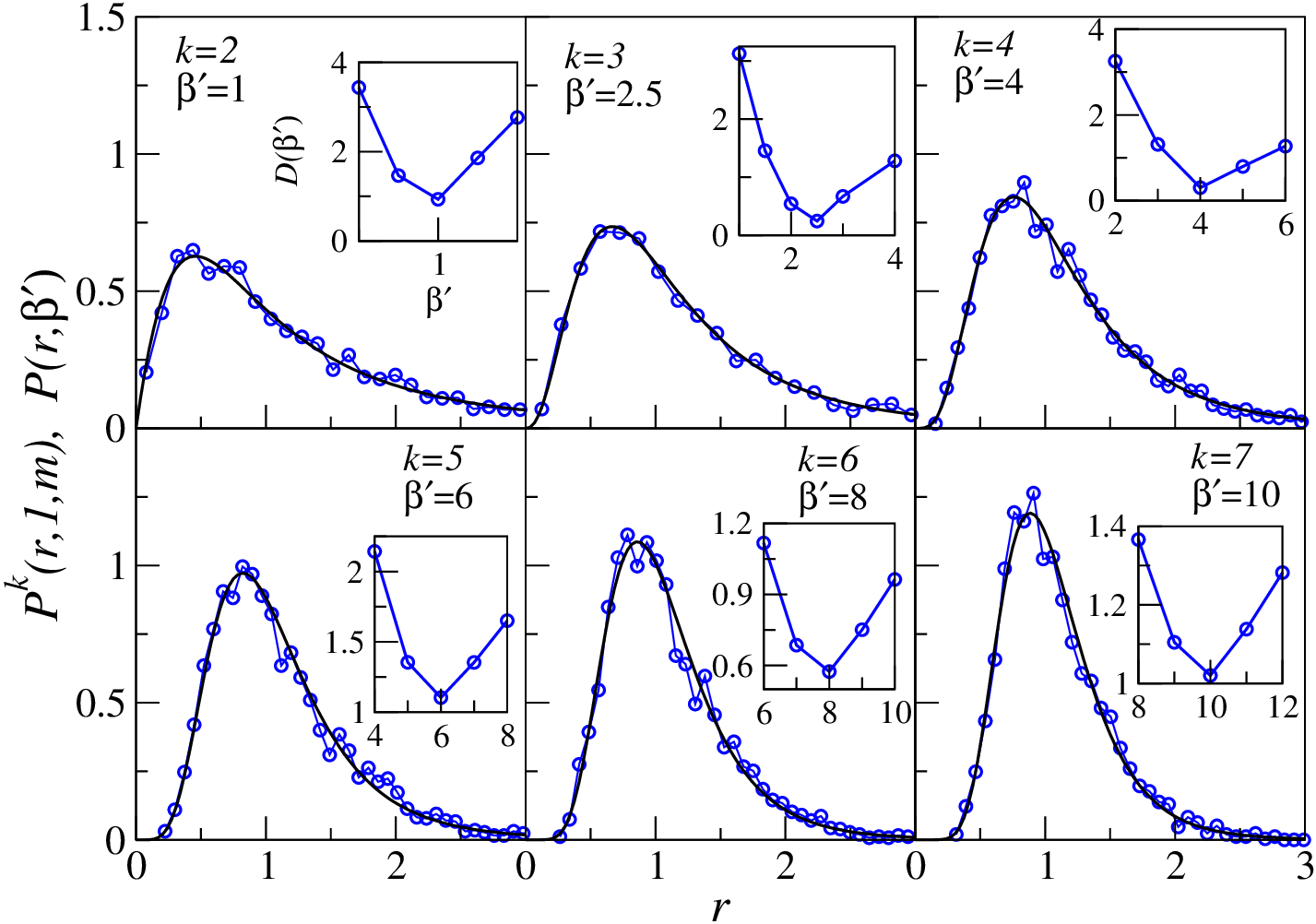}
\caption{(Color online) Same as Fig.~\ref{fig:SpinChain2Sym1} but for $L=14$ and $\beta'=1,2.5,4,6,8,10$.
Here $7$ spins are up and the dimension of the block-matrix is $3432$.}
\label{fig:SpinChain4Sym1}
\end{center}
\end{figure}
\begin{figure}[h!]
\begin{center}
\includegraphics*[scale=0.35]{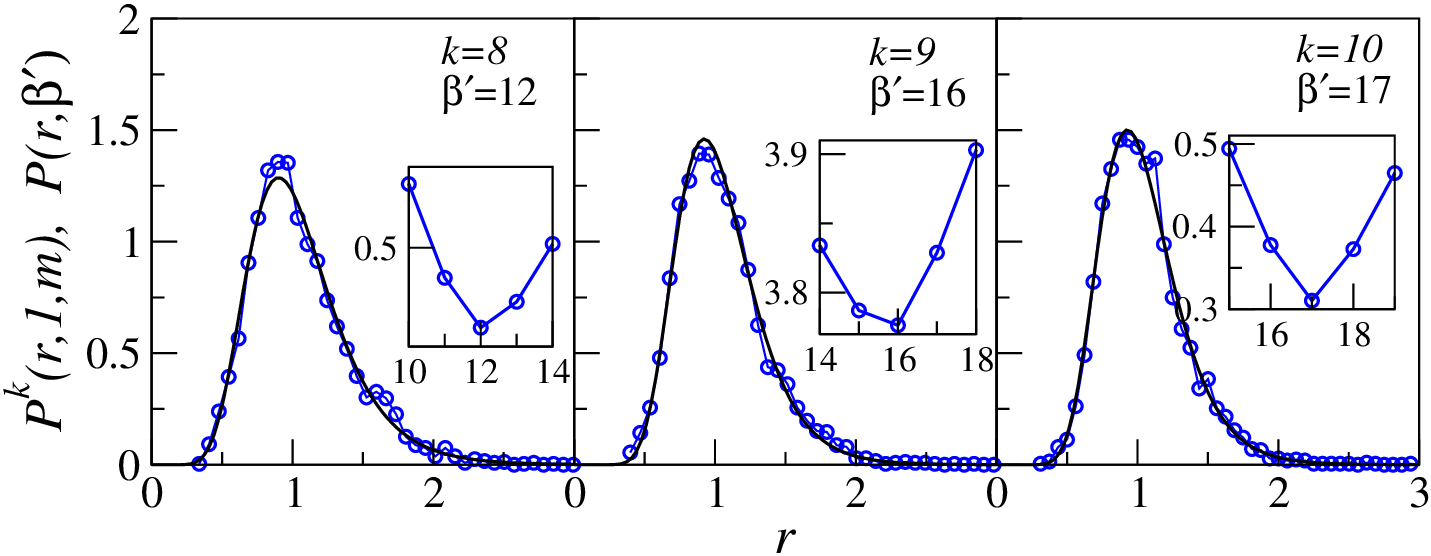}
\caption{(Color online) Same as Fig.~\ref{fig:SpinChain4Sym1} but for $k=8$, $9$, $10$ and $\beta'=12$, $16$, $17$.}
\label{fig:SpinChain4Sym2}
\end{center}
\end{figure}


\section{Superposition and higher-order spacing ratios in CUE}
\label{SuperpositionCUE}

In this section, we study higher-order spacings ratio in the superposition of CUEs on the lines of 
Section~\ref{sec:SuperpositionCOE} where superposition of COEs is studied. The CUE is used in modelling Hamiltonians which lack 
time-reversal symmetry \cite{Mehtabook}. The results are tabulated in Table~\ref{CUEtable1} for $m=2$ to $5$ and various values of 
$k$. In this case also except few all values of $\beta'$ are the whole number.
The results are plotted in Figs.\ref{fig:CUE3added1}, \ref{fig:CUE3added2} and \ref{fig:CUE3added3}.
The Fig.\ref{fig:CUE3added3} shows noninteger values of $\beta'$ which is found by best fit.

\begin{table}[t!]
\begin{center}
\begin{tabular}{|c|c|c|c|c|}
\hline
$k$ & $m=2$ & $m=3$ & $m=4$ & $m=5$\\
&$\beta'$&$\beta'$&$\beta'$&$\beta'$\\
\hline
1&0&0&0&0\\
2&3&1.5&1&0.75\\
3&6&4&3&2\\
4&11&7&6&4\\
5&15&11&8&7\\
6&22&15&11&10\\
7&28&20&15&13\\
8&36&25&20&16\\
9&43&31&24&20\\
10&55&37&29&24\\
11&64&44&34&28\\
12&75&51&40&33\\
13&&59&&\\
14&&67&&\\
15&&76&&\\
16&&85&&\\
\hline
\end{tabular}
\caption{Tabulation of higher-order indices $\beta'$ for various $k$ and superposition of $m$ CUEs, each 
having dimension $N=10000$.}
\label{CUEtable1}
\end{center}
\end{table}

  
In the case of superposing of CUEs $m=3$ is an interesting case for which we have obtained a scaling relation 
for even and odd values of $k$ and $\beta'$ as given below:
\begin{equation}
\label{Eq:CUEseries1}
 \beta'= 4k- 9 + \frac{(k-4)(k-6)}{4},\,\,\,k=4,6,8,\ldots
\end{equation}
and
\begin{equation}
\label{Eq:CUEseries2}
 \beta'=1+k+\frac{5(k-3)}{2}+ \frac{(k-3)(k-5)}{4},\,\,\,k=3,5,7,\ldots
\end{equation}
By suitable change of variables as $l=k/2$ the Eq.~\ref{Eq:CUEseries1} reduces to $\beta'=l^2+3l-3$ where 
$l=1,2,3,\ldots$. While using $q=(k-1)/2$ the Eq.~\ref{Eq:CUEseries2} reduces to $\beta'=q^2+4q-1$ where 
$q=1,2,3\ldots$.
Comparing these series with Eq.~(\ref{Eq:HigherOrder}) one can see that it does not reduce to 
Eq.~(\ref{Eq:HigherOrder}) for any $\beta$. Thus, using the scaling relations in
Eqs.~(\ref{Eq:CUEseries1}) and (\ref{Eq:CUEseries2}) no statement can be made at the level of jpdf of 
circular $\beta$-ensemble.

 

\begin{figure}[h!]
\begin{center}
\includegraphics*[scale=0.35]{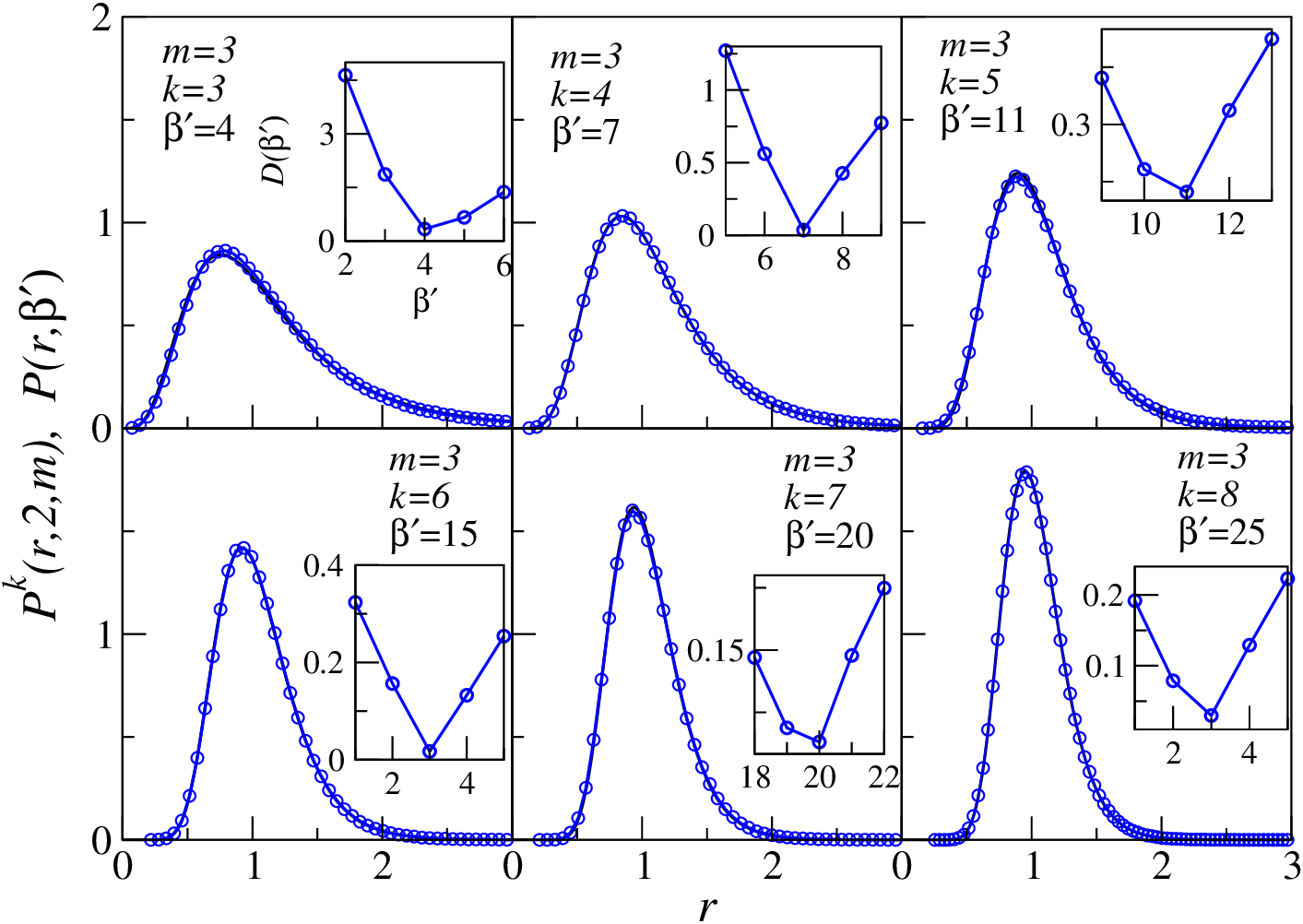}
\caption{(Color online) Distribution of the $k$-th ($3$ to $8$) order spacing ratios (circles) for a superposition of 
$m=3$ CUE spectra. The dimension of the matrices is $N=10000$. The solid curve corresponds to 
$P(r,\beta')$ as given in Eq.(\ref{Eq:PRBeta}) with $\beta'$ given in Table~\ref{CUEtable1}. 
The insets shows $D$ as a function of $\beta'$.}
\label{fig:CUE3added1}
\end{center}
\end{figure}
\begin{figure}[h!]
\begin{center}
\includegraphics*[scale=0.35]{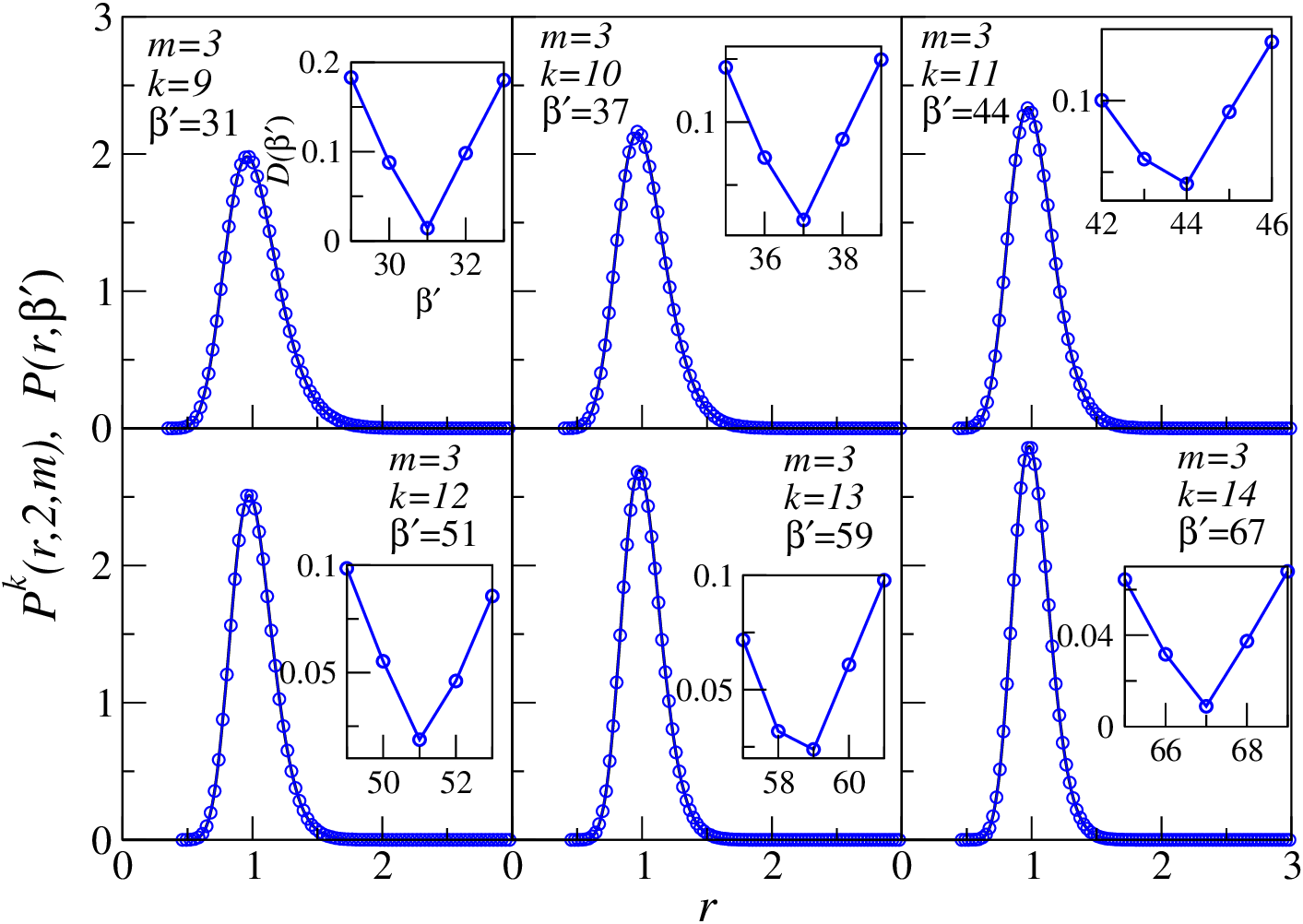}
\caption{(Color online) Same as Fig.~\ref{fig:CUE3added1} but for $k=9$ to $14$ with corresponding 
$\beta'$ given in Table~\ref{CUEtable1}.}
\label{fig:CUE3added2}
\end{center}
\end{figure}
\begin{figure}[h!]
\begin{center}
\includegraphics*[scale=0.35]{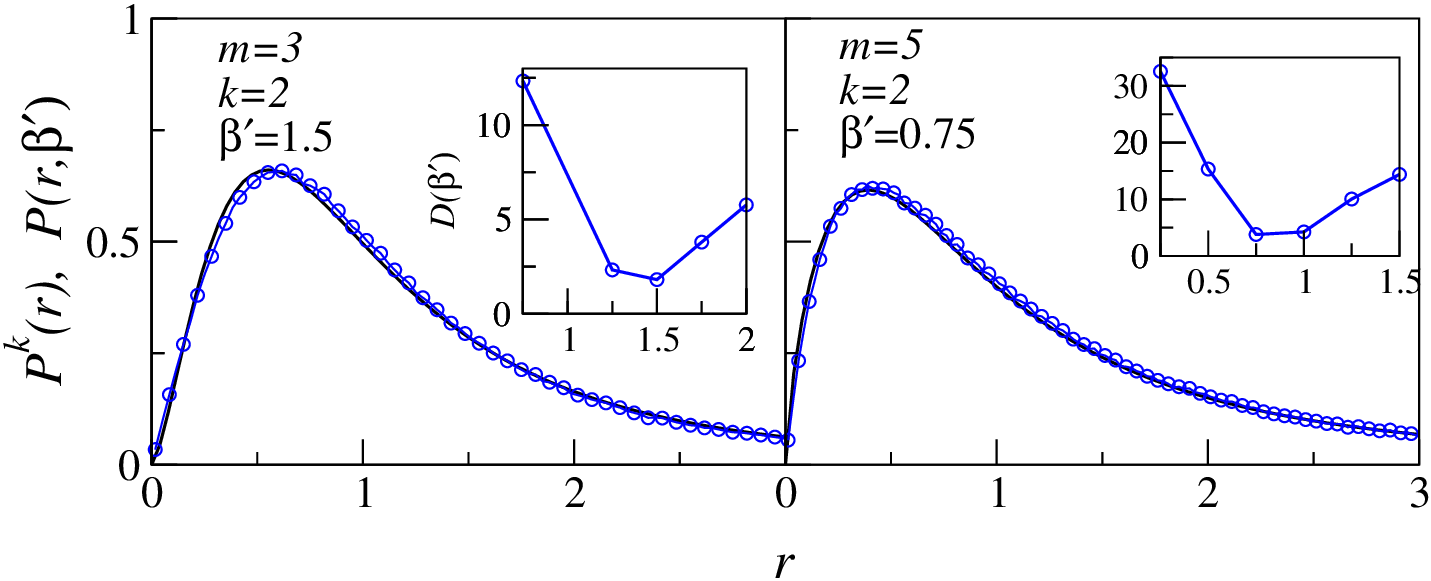}
\caption{(Color online) Same as Fig.~\ref{fig:CUE3added1} but for $k=2$, $m=3$ and $5$  with corresponding noninteger values 
of $\beta'$ given in Table~\ref{CUEtable1}.}
\label{fig:CUE3added3}
\end{center}
\end{figure}

\section{Superposition and higher-order spacing ratios in CSE}
\label{Sec:SuperpositionCSE}

In this section, we study higher-order spacings ratio in the superposition of CSEs on the lines of 
Section~\ref{sec:SuperpositionCOE}. The CSE is used in modeling Hamiltonians with time-reversal symmetry and 
half-integer spin interaction \cite{Mehtabook,zyczkowski1995random}. The results are tabulated in Table~\ref{CSEtable1} for 
$m=2$ to $6$ and various values of $k$. The results are plotted in Fig.~\ref{fig:CSEadded1} and \ref{fig:CSEadded2}.
In this case also except few all values of $\beta'$ are whole number.

Based on results in Table~\ref{CSEtable1}, we have given two conjectures at the level of spectral fluctuations. 
The first conjecture is as follows:
\begin{equation}\label{Eq:Conjecture4}
P^{k}(r,4,m)=P(r,\beta'), \; \mbox{for} \; \beta'=2k+1=2m+3
\end{equation}
and $m\geq2$, while the second one is as follows:
\begin{equation}\label{Eq:Conjecture5}
P^{k}(r,4,m)=P(r,\beta'), \; \mbox{for} \; \beta'=2(k-1)=2(m-2)
\end{equation}
and $m\geq3$. Our conjectures hold true for asymptotic value of $N$. 

Although we are able to find scaling relations only for few cases, but for given $m$ one can compare the 
sequence of $\beta'$ as a function of $k$ with that of $m'\neq m$, within and across the Tables~\ref{COEtable1}, 
\ref{CUEtable1} and \ref{CSEtable1}. It can be seen that these sequences are unique for given $m$ and the type 
of ensemble considered. One can also see that this is an increasing sequence on the lines of earlier result in
Eq.\ref{Eq:HigherOrder} from Refs.\cite{Harshini2018a,UdaysinhBhosaleScaling2018}.
With this observation, we would like to conjecture that for a given number of symmetries $m$ and the Dyson index 
$\beta$ of the circular ensemble or a quantum chaotic system, the sequence of $\beta'$ is an increasing function 
of $k$ and completely characterize the ensemble or the system uniquely.

Similarly, for given $k$ one can compare the sequence of $\beta'$ as a function of $m$ with that of $k'\neq k$, 
within and across the Tables~\ref{COEtable1}, \ref{CUEtable1} and \ref{CSEtable1}. Also, the sequences are unique 
for given $k$ and decreasing. Thus, with this observation, we would like to conjecture that for given $k$
and the Dyson index $\beta$, the sequence of $\beta'$ is decreasing as a function of $m$ and is unique. 
This can be interpreted physically as follows: The level repulsion present in the eigenvalues characterized by 
$\beta'$  for given $m$ and $k$ is reduced as $m$ is increased and is reflected in the reduction of the new value of $\beta'$. The implication of this conjecture is that for given $\beta$, $k$ and $m\rightarrow\infty$ we will see
$\beta'\rightarrow 0$. It can be seen easily that one conjecture does not imply the second one.


Proving our results mathematically is challenging but we give an intuitive argument for the last conjecture.
We know that NN as well as the $k$-th eigenvalues in circular or Gaussian spectra repel each other 
\cite{Forresterbook}. It has also been established mathematically (analytical and numerical) that when two 
same dimensional COE/GOE spectra are superposed the NN don't repel each other (level clustering) in the limit 
of matrix dimensions tending to infinity, which results in their spacings distribution to be Poissonian
\cite{Mehtabook,Guhr98,KarolTensor2012,KarolExtremal2013,tkocz2013note}.
In other words, before superposition, the level repulsion present (characterized by $\beta=1$) has now vanished 
after superposition (characterized by $\beta=0$). And this is what is also observed in our last conjecture
i.e. for given $k$ and $m$ the value of $\beta'$ (characterizing the repulsion between the $k$-th eigenvalues) 
reduces as $m$ is increased. In most of the cases we studied here, although not guaranteed, these reductions 
are integer number.
The special case of our conjecture where $k=1$, $N=2$ and $m\rightarrow\infty$ is shown
to have NN spacing distribution as Poissonian in Ref.\cite{KarolTensor2012}.



\begin{figure}[t!]
\begin{center}
\includegraphics*[scale=0.35]{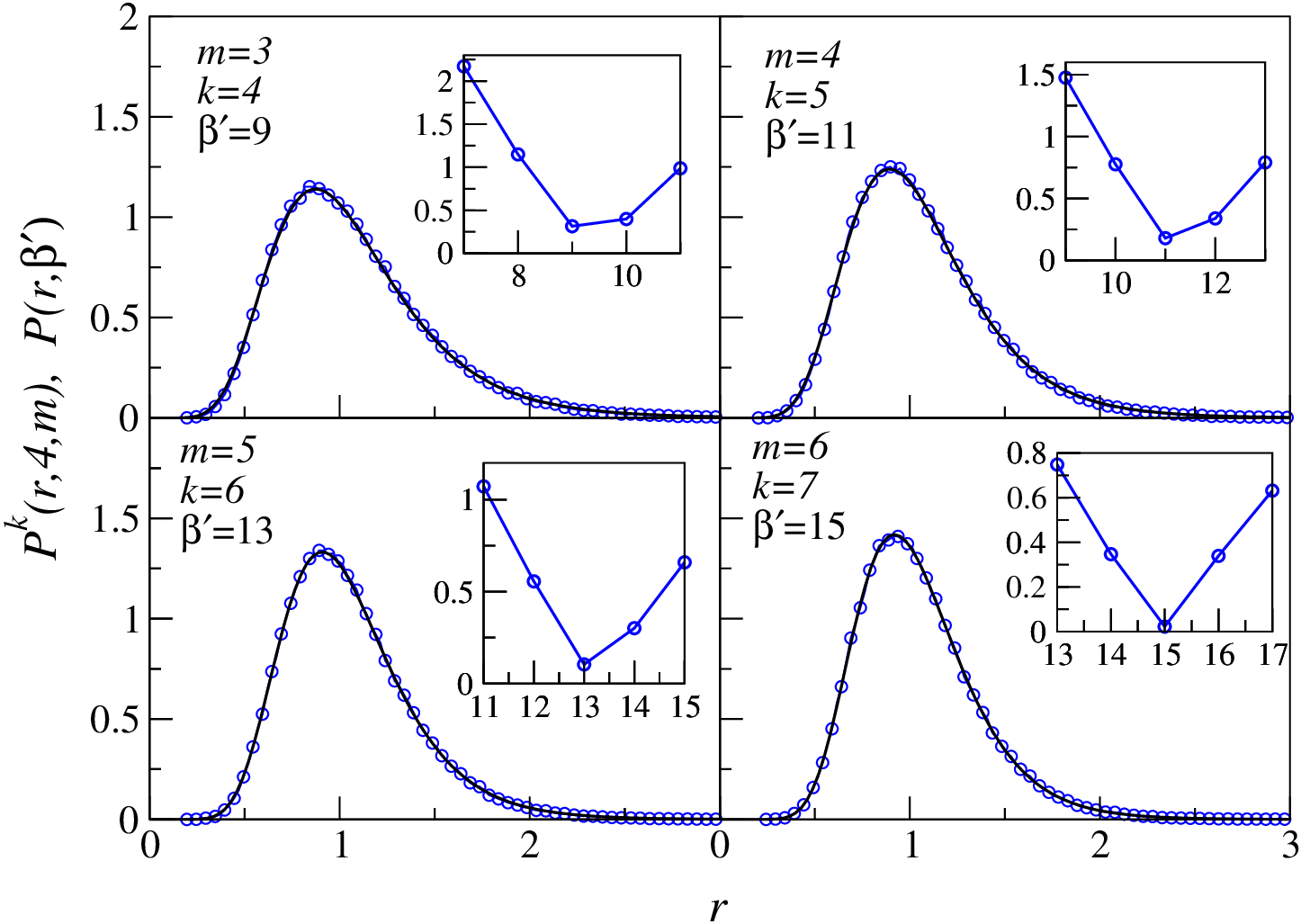}
\caption{(Color online) Distribution of the $k$-th order spacing ratios (circles) for various values of $k$ 
and superposition of $m$ CSE spectra. The dimension of the matrices is $N=1000$. The solid curve corresponds to 
$P(r,\beta')$ as given in Eq.(\ref{Eq:PRBeta}) with $\beta'$ given in Table~\ref{CSEtable1}. The insets shows 
$D$ as a function of $\beta'$.}
\label{fig:CSEadded1}
\end{center}
\end{figure}

\begin{figure}[t!]
\begin{center}
\includegraphics*[scale=0.35]{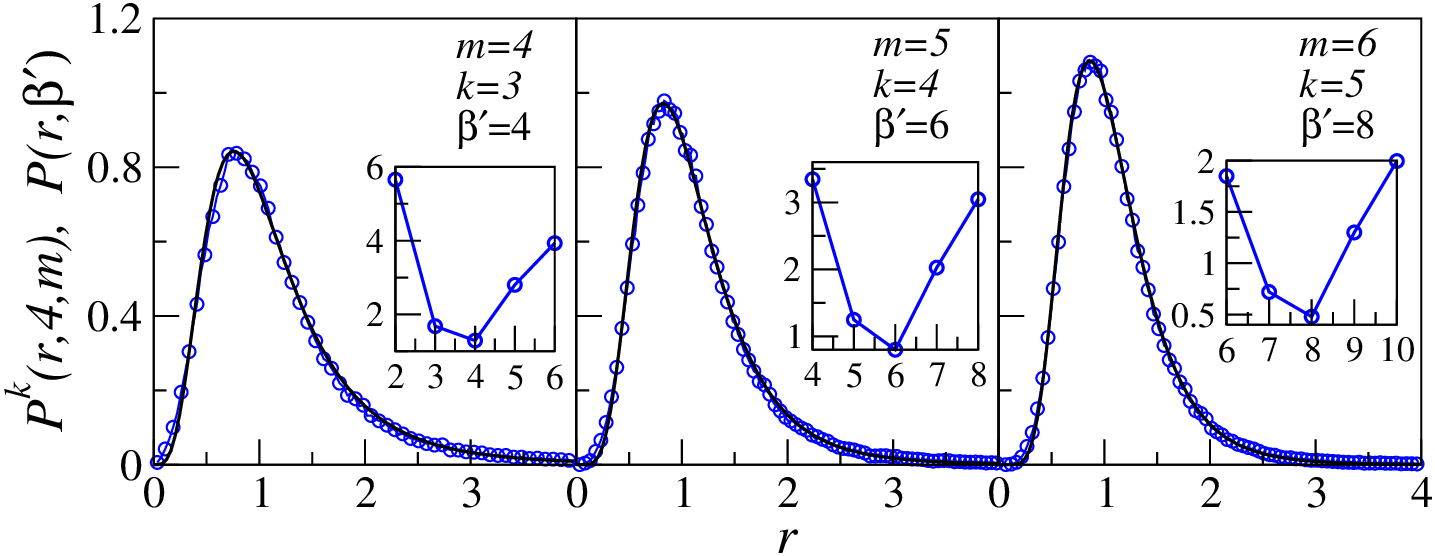}
\caption{(Color online) Same as Fig.~\ref{fig:CSEadded1} but for different values of $k$ and $m$.}
\label{fig:CSEadded2}
\end{center}
\end{figure}

\begin{table}[h!]
\begin{center}
\begin{tabular}{|c|c|c|c|c|c|c|}
\hline 
$k$ & $m=2$ & $m=3$ & $m=4$ & $m=5$ & $m=6$ \\
&$\beta'$&$\beta'$&$\beta'$&$\beta'$&$\beta'$ \\ 
\hline
1&0&0&0&0&0\\
2&5&2&1&0.75&0.75\\
3&7&7&4&2.5&2\\
4&18&9&8 &6&4\\
5&21&14&11&10&8\\
6&36&23 &14&13&12\\
7&39&27 &21&16&15\\
8&60&35 &28&20&18\\
9&63&47 &33&27&21\\
10&88&53&38&35&27\\
11&92&63&48&40&34\\
12&122&79&59&44&40\\
\hline
\end{tabular}
\caption{Tabulation of higher-order indices $\beta'$ for various $k$ and superposition of $m$ CSEs each 
having dimension $N=1000$.}
\label{CSEtable1}
\end{center}
\end{table}

\section{Numerical methods} 
\label{Sec:NumericalMethods}


Now, various numerical pieces of evidence supporting our results are presented. These best fits are checked with the 
numerical data quantitatively. 

As a numerical check for our claims, analysis using Eq.~(\ref{Eq:HigherOrder}) for $P(r,\beta')$ is 
carried, where no fitting parameter is involved. For this, the difference between the cumulative distributions 
is numerically found and defined as follows:
\begin{equation}\label{Eq:DBetaFormula}
D(\beta')= \sum_i \,|F^{k}_{\mbox{obs}}(r_i,\beta,m)-F(r_i,\beta')|,
\end{equation}
where $F^{k}_{\mbox{obs}}(r,\beta,m)$ and $F(r,\beta')$ denotes cumulative distribution functions corresponding 
to the observed histogram $P^{k}_{\mbox{obs}}(r,1,m)$ and the numerical fit or the postulated function 
$P(r,\beta')$ respectively. This definition has been used in earlier works 
\cite{tekur2018symmetry,UdaysinhBhosaleScaling2018,Harshini2018a} in similar kind of analysis. It can be seen 
that, $D(\beta')$ can take any positive value (upper bound) depending on the range of $i$ in the summation, but 
is minimum only for that value of $\beta'$ for which $P(r,\beta')$ is best fit for the observed histogram. The 
values of $k$ for given $m$ are same as that in Figs.~\ref{fig:COE2added}, \ref{fig:COE3added}, \ref{fig:COE4added}, 
\ref{fig:COE5added}, \ref{fig:COE6added}, \ref{fig:Kickedtop1}, \ref{fig:Kickedtop2}, \ref{fig:Nucleardata1}, \ref{fig:SpinChain2Sym1},
\ref{fig:SpinChain4Sym1}, \ref{fig:SpinChain4Sym2}, \ref{fig:CUE3added1}, \ref{fig:CUE3added2}, \ref{fig:CSEadded1} and 
\ref{fig:CSEadded2}. The results of $D(\beta')$ are shown in 
the insets of these figures. It can be seen that the minima of $D(\beta')$ in each case coincides remarkably 
with that of corresponding $\beta'$ from the main figures.  

After finding the best fit for the observed data using $D(\beta')$, we go on to check how close the 
two probability distributions and their respective cumulative fuctions are. Firstly, the overlap ($p$) 
between the probability plots in 
%
Figs.~\ref{fig:COE2added}, \ref{fig:COE3added}, \ref{fig:COE4added}, 
\ref{fig:COE5added}, \ref{fig:COE6added}, \ref{fig:Kickedtop1}, \ref{fig:Kickedtop2}, \ref{fig:Nucleardata1}, \ref{fig:SpinChain2Sym1},
\ref{fig:SpinChain4Sym1}, \ref{fig:SpinChain4Sym2}, \ref{fig:CUE3added1}, \ref{fig:CUE3added2}, \ref{fig:CSEadded1} and 
\ref{fig:CSEadded2} is calculated using the following definition:  
\begin{equation}
 p=1-\int |P^{k}_{\mbox{obs}}(r,\beta,m)-P(r,\beta')|\,dr.
\end{equation} 
Secondly, the cumulative distribution functions corresponding to observed data 
$P^{k}_{\mbox{obs}}(r,\beta,m)$ and $P(r,\beta')$ is studied. The maximum absolute difference ($d$) between 
these cumulative distributions is calculated using the following definition: 
\begin{equation}
d= \underset{r_i}{\mbox{Sup}}\,|F^{k}_{\mbox{obs}}(r_i,\beta,m)-F(r_i,\beta')|.
\end{equation}
By definition $0\leq p,d\leq 1$ and larger (smaller) 
value of $p$ ($d$) will indicate that the numerically observed distribution is close to that of the postulated 
one. Unlike $D(\beta')$, these values will only improve as the range of $i$ is increased. The values are shown in 
Tables~\ref{Kolmogorov1}, \ref{Kolmogorov2}, \ref{Kolmogorov3}, \ref{Kolmogorov6}, \ref{Kolmogorov4}, \ref{Kolmogorov5} and 
\ref{Kolmogorov7}. The Tables~\ref{Kolmogorov1}, \ref{Kolmogorov2}, \ref{Kolmogorov3} and \ref{Kolmogorov6} gives strong evidences 
for our results in Tables~\ref{COEtable1}, \ref{CUEtable1} and \ref{CSEtable1} corresponding to the superposition of COE, CUE and CSE 
respectively. In Tables~\ref{Kolmogorov4}, \ref{Kolmogorov5} and \ref{Kolmogorov7} results are shown for the physical system
of QKT, measured nuclear resonances and the spin Hamiltonian as plotted in Figs.\ref{fig:Kickedtop1}, \ref{fig:Kickedtop2}, 
\ref{fig:Nucleardata1}, \ref{fig:SpinChain2Sym1}, \ref{fig:SpinChain4Sym1} and \ref{fig:SpinChain4Sym2}. It can be seen that in all 
the cases the values of $p$ and $d$ shown in these tables gives strong evidences for our results in these figures. The effect of 
small sample size in the case of nuclear resonances can be seen in the Table~\ref{Kolmogorov5}.

 
\begin{table}[h!]
\begin{center}
\begin{tabular}{|c|c|c|c|c|c|c|c|c|c|c|}
\hline 
 $k$&$m=2$ & &$k$&$m=3$ &  & $k$ & $m=4$   \\
& $p,d$ &&&$p,d$&& &$p,d$\\ 
\hline
3& $0.983,0.004153$ && 3& $0.992,0.00085$ &&2  &$0.991,0.00164$ \\
5& $0.986,0.003421$ && 4& $0.991,0.00117$ &&4  &$0.991,0.00156$ \\
7& $0.989,0.002506$&& 6& $0.994,0.00057$ &&5  &$0.992,0.00116$ \\
9& $0.994,0.001173$ && 7& $0.989,0.00246$ &&6  &$0.995,0.00081$ \\
11&$0.996,0.000651$ && 8& $0.988,0.00268$ &&7  &$0.988,0.003263$ \\
13&$0.993,0.001836$ && 9& $0.994,0.00105$ &&8  &$0.995,0.00079$\\
& && 10& $0.988,0.0023$ &&9  &$0.986,0.003078$ \\
& && 11& $0.995,0.0014$ &&11 &$0.996,0.000431$ \\
& && 12&$0.995,0.00077$ &&13 &$0.994,0.001768$ \\
& && 13&$0.994,0.00077$ &&15 &$0.996,0.000466$ \\
\hline
\end{tabular}
\caption{The overlap probability $p$ and the maximum absolute difference $d$ for the results on the superposition
of COEs for $m=2,3,4$ and various $k$'s.}
\label{Kolmogorov1}
\end{center}
\end{table}
\begin{table}[h!]
\begin{center}
\begin{tabular}{|c|c|c|c|c|c|c|c|c|c|c|}
\hline 
$k$& $m=5$  &&$k$& $m=6$ &  & $k$ & $m=7$   \\
& $p,d$ &&&$p,d$ &&&$p,d$\\ 
\hline
2& $0.957,0.01088$& &3& $0.967,0.00822$ && 3&$0.941,0.001464$\\
3& $0.989,0.00186$& &4& $0.991,0.0009$  && 4&$0.978,0.005134$\\
6& $0.995,0.00044$& &5& $0.965,0.00882$ && 5&$0.991,0.000112$\\
7& $0.987,0.00253$& &6& $0.976,0.00562$ && 6&$0.978,0.005358$\\
8& $0.981,0.00479$& &7& $0.988,0.00248$ && 7&$0.974,0.006416$\\
9& $0.996,0.00046$& &8& $0.996,0.00046$ && 8&$0.987,0.003019$\\
10&$0.990,0.00232$& &9& $0.983,0.00416$ && 9&$0.990,0.002234$\\
11&$0.990,0.00264$& &10&$0.991,0.00241$ &&10&$0.995,0.00096$\\
12&$0.996,0.00206$& &12&$0.996,0.00054$ &&15&$0.997,0.00094$\\
13&$0.996,0.00037$& &16&$0.996,0.00138$ &&20&$0.996,0.0013$\\
\hline
\end{tabular}
\caption{Same as Table \ref{Kolmogorov1} but for $m=5,6$ and $7$.}
\label{Kolmogorov2}
\end{center}
\end{table}  
\begin{table}[h!]
\begin{center}
\begin{tabular}{|c|c|c|c|c|c|c|c|c|c|c|}
\hline 
$k$&$m=3$ &  & $k$ & $m=3$   \\
& $p,d$ &&&$p,d$\\ 
\hline
3&$0.9673,0.008161$  && 9&$0.9977,0.0006759$\\
4&$0.9965,0.0007662$ &&10&$0.9976,0.00080449$\\
5&$0.9813,0.005530$  &&11&$0.9965,0.0015069$\\
6&$0.9948,0.000936$  &&12&$0.9961,0.0005231$\\
7&$0.9893,0.004329$  &&13&$0.9971,0.0011679$\\
8&$0.9960,0.0017572$ &&14&$0.99709,0.0006004$\\
\hline
\end{tabular}
\caption{Same as Table \ref{Kolmogorov1} but for CUE and $m=3$.}
\label{Kolmogorov3}
\end{center}
\end{table} 
\begin{table}[h!]
\begin{center}
\begin{tabular}{|c|c|c|c|c|c|c|c|c|c|c|c|c|}
\hline 
$k$&$m=4$ &  &$k$ & $m=5$ && $k$ & $m=6$  \\
& $p,d$ &&&$p,d$ &&& $p,d$ \\ 
\hline
2&$0.9821,0.00884$  && 4&$0.9966,0.00657$ && 3& $0.9698,0.00709$\\
3&$0.9538,0.009765$  && 5&$0.9896,0.00114$&& 4& $0.9530,0.00870$\\
4&$0.9784,0.005174$ &&6&$0.9896,0.00216$&& 5& $0.9760,0.00540$\\
5&$0.9856,0.00353$  &&7&$0.9880,0.00209$&& 6& $ 0.9850,0.00379$\\
6&$0.988,0.002077$  &&8&$0.9888,0.00247$&& 7& $0.9918,0.00129$\\
7&$0.9907,0.002545$  &&9&$0.9872,0.00272$&& 8& $0.9874,0.00298$\\
8&$0.9873,0.002881$ &&10&$0.9948,0.00132$&& 9& $0.9931,0.00100$\\
9&$0.9913,0.002375$  && 11&$0.9941,0.00114$&& 10& $ 0.9916,0.00195$\\
10&$0.9917,0.001246$  && 12&$0.9923,0.00152$&& 11& $ 0.9907,0.00179$\\
11&$0.9938,0.001067$  && &   && 12& $0.9897,0.00212$\\
12&$0.9636,0.001533$  && &    && &\\
\hline
\end{tabular}
\caption{Same as Table \ref{Kolmogorov1} but for CSE and $m=4$, $5$ and $6$.}
\label{Kolmogorov6}
\end{center}
\end{table} 
\begin{table}[h!]
\begin{center}
\begin{tabular}{|c|c|c|c|c|c|c|c|c|c|c|}
\hline 
$k$&$\beta'$ &$p,d$   &  & $k$ & $\beta'$& $p,d$   \\
\hline
2&2&$0.97263,0.003742$  &&8&23&$0.97814,0.002949$\\
3&4&$0.95997,0.008608$  &&9&28&$0.98059,0.002558$\\
4&7&$0.97056,0.003950$  &&10&34&$0.97707,0.002847$\\
5&10&$0.97393,0.004558$ &&11&40&$0.97981,0.001904$\\
6&14&$0.98118,0.001842$ &&12&47&$0.97895,0.004052$\\
7&18&$0.97883,0.003812$ &&13&54&$0.98250,0.002808$\\
\hline
\end{tabular}
\caption{The overlap probability $p$ and the maximum absolute difference $d$ for the distribution of 
higher-order spacing ratios using the eigenvalues of QKT in chaotic case. The value of $\beta'$ for given $k$
is same as that in Fig.~\ref{fig:Kickedtop1} and \ref{fig:Kickedtop2}.}
\label{Kolmogorov4}
\end{center}
\end{table} 
\begin{table}[h!]
\begin{center}
\begin{tabular}{|c|c|c|c|c|c|c|c|c|c|c|}
\hline 
$k$&$\beta'$ &$p,d$   &  & $k$ & $\beta'$& $p,d$   \\
\hline
2&2&$0.88868,0.018197$  &&4&6&$0.89450,0.028502$\\
3&4&$0.85402,0.027265$  &&5&8&$0.87177,0.031754$\\
\hline
\end{tabular}
\caption{Same as Table~\ref{Kolmogorov4} but for the data of nuclear resonances. The value of $\beta'$ for 
given $k$ is same as that in Fig.~\ref{fig:Nucleardata1}.}
\label{Kolmogorov5}
\end{center}
\end{table} 

\begin{table}[h!]
\begin{center}
\begin{tabular}{|c|c|c|c|c|c|c|c|c|c|c|}
\hline 
$k$&$\beta'$ &$p,d$   &  & $k$ & $\beta'$& $p,d$   \\
\hline
2&2&$0.93226,0.01725$  &&2&1&$0.92570,0.01059$\\
3&4&$0.92965,0.01293$  &&3&2.5&$0.95498,0.01021$\\
4&7&$0.93976,0.01779$  &&4&4&$0.95163,0.00550$\\
5&11&$0.92546,0.03858$ &&5&6&$0.93557,0.01493$\\
6&12&$0.95894,0.00826$ &&6&8&$0.93278,0.01389$\\
7&16&$0.94367,0.01096$ &&7&10&$0.94717,0.01175$\\
&&                    &&8&12&$0.94094,0.00822$\\
&&                    &&9&16&$0.94468,0.01891$\\
&&                    &&10&17&$0.94992,0.00791$\\
\hline
\end{tabular}
\caption{The overlap probability $p$ and the maximum absolute difference $d$ for the distribution of 
higher-order spacing ratios using the eigenvalues of spin chains in chaotic case. 
The value of $\beta'$ for given $k$ is same as that in Figs.~\ref{fig:SpinChain2Sym1} (left), \ref{fig:SpinChain4Sym1} (right)
and \ref{fig:SpinChain4Sym2} (right).}
\label{Kolmogorov7}
\end{center}
\end{table}

%

 
 

 


\section{SUMMARY AND CONCLUSIONS}
\label{Sec:Summary}

This paper has studied the long-range correlations in the superposed spectra of COE, CUE and CSE using 
higher-order spacing ratios. We have given a table for the modified Dyson indices ($\beta'$) corresponding to the 
distribution of $k$-th order spacing ratio when $m$ number of matrices each from COEs, CUEs and CSEs are 
superposed. For the case when two 
COEs are superposed two scaling relations relating $\beta'$ and $k$ are found for even and odd values of $k$
respectively. 
The relation corresponding to even $k$ is related to the earlier result on the connection between CUE and 
superposition of two COEs at the level of jpdf \cite{Dyson623,gunson1962proof}. 
Conjectures on the lines of Ref.\cite{tekur2018symmetry} are given.
For the case of COE, it is conjectured that for given $m$, the distribution of the $k$-th order spacing ratio 
is related to $\beta'$ such that the relation $\beta'=k+1=m+2$ for $m\geq2$ and $\beta'=k-1=m-4$ for $m \geq 5$
holds true. Similarly, for the case of CSE, the relation $\beta'=2k+1=2m+3$ for $m\geq2$ and
$\beta'=2(k-1)=2(m-2)$ for $m \geq 3$ holds true. 

 
We have tested our results on three different physical systems.
The first one is the QKT in the quantum chaotic limit belonging to COE.
The other two are the measured nuclear resonances and a spin Hamiltonian both corresponding to the GOE.
%
These systems are known to have symmetries. 
For the case of QKT, we have tested our RMT results of $m=2$ case up to $k=13$ and found very good agreement.
This agrees with the earlier analytical result from the Ref.\cite{Hakke87}, where it is shown 
that its Hamiltonian has two symmetries.
Whereas in the case of nuclear resonances we could find agreement only up to $k=3$ due to nonuniform density as well as small sample 
size. But the results of $k=2$ and $3$ were enough to conclude the presence of two symmetries using the uniqueness of our tabulated 
COE results. The third system we considered is the quantum chaotic spin Hamiltonian. Depending on the values of the parameter a given spin 
sector can have two or four symmetries. In both the cases we tested our COE results of $m=2$ and $m=4$. In this case, also we observed 
effects of nonuniform density but at large values of $k$ compared to the previous case due to the large matrix dimension. These results 
imply that our RMT results hold very well to quantum chaotic physical systems modeled by circular ensembles. For others, we may see 
the deviations for higher values of $k$ depending on the system and its matrix dimension or the sample size. 
Despite this we could successfully find the symmetries of systems modeled by Gaussian ensembles.
Looking at our results and the 
Refs.\cite{Mehtabook,Forresterbook,tekur2018symmetry,tekurhigher2018,porter1963further,kahn1963statistical} specially the 
Ref.\cite{tekurhigher2018} our results can be claimed to be true for Gaussian ensembles in the limit of large matrix dimensions. 
Thus, our results can be used to find the symmetries in unknown physical systems. 
 
 



For the case of $m=3$ superposition of CUEs two scaling relations relating $\beta'$ and $k$ are found for 
even and odd values of $k$. These scaling relations along with other results are confirmed numerically using 
large matrix dimensions. We have used various numerical tests for the verification of our results. 
We conjectured that for given $m$ ($k$), the sequence of $\beta'$ as a function of $k$ ($m$) is 
increasing (decreasing) and is unique to a given circular $\beta$-ensemble. 
%
%
As a corollary, finding symmetries as well as whether a given quantum chaotic system is time-reversal invariant 
(with or without the spin degree of freedom) or not can be found unambiguously. 
%
%
%
The Gaussian ensembles have been implemented in various 
experimental systems \cite{GuhrCorrelation1997,RichterGUE2000,SeligmanTransition2002,RichterSpectral2014,StockmannGSERealization2016}.
Thus, our circular ensemble results can be tested using these experiments by taking experimental systems
with suitable geometrical symmetry corresponding to given $m$.

This work has given rise to new future directions as well. We would like to test our results as an additional 
and stringent test for finding symmetries in various other quantum complex systems 
\cite{tekur2018symmetry,giraud2020probing,ProsenQFT2021}. Various quantum chaotic systems with and without time-reversal invariance 
and having additional symmetries can be tested. Our study can be extended to the case when matrices of 
unequal dimensions are superposed which will be relevant to understand symmetries in various other spin systems 
\cite{binder1986spin,amico08,tekur2018symmetry,giraud2020probing}. Our study can be extended to other relevant ensembles from RMT, for 
example the ensembles with chiral symmetry 
\cite{fyodorov2002correlation,damgaard2011chiral,kaymak2014supersymmetry,BeenakkerMajorana2015,akemann2017random,PragyaShuklaChiral2020,RichterMicrowave2020}
and Wishart ensemble \cite{Wishart28,Satyalargeright,akemann2011oxford,Fridman12,UdaysinhBhosaleScaling2018}.


\section{Acknowledgments}
Authors are grateful to M. S. Santhanam, Harshini Tekur and Ravi Prakash for useful comments and discussions at 
various stages of this paper. UTB thanks Rukmani Bai and Hrushikesh Sable for helping in the numerical data. 
The results presented in the paper are based on the computations using {\it Mathematica 9} in Vikram-100, 
the 100TFLOP HPC Cluster at Physical Research Laboratory, Ahmedabad, India. 

 
 

\bibliography{reference22013,reference22,reference221}
\end{document}